\documentclass[conference]{IEEEtran}
\IEEEoverridecommandlockouts
\usepackage{cite}
\usepackage{amsmath,amssymb,amsfonts}
\usepackage{algorithmic}
\usepackage{graphicx}
\usepackage{textcomp}
\usepackage{xcolor}
\usepackage[labelsep=period]{caption}
\usepackage{subcaption} 
\usepackage{array}

\newcommand{\perth}{\texttt{ibm\_perth}}   
\newcommand{\brisbane}{\texttt{ibm\_brisbane}} 
\newcommand{\twoq}{2Q} 
\newcommand{\centered}[1]{\begin{tabular}{l} #1 \end{tabular}}

\def\BibTeX{{\rm B\kern-.05em{\sc i\kern-.025em b}\kern-.08em
    T\kern-.1667em\lower.7ex\hbox{E}\kern-.125emX}}
\begin{document}


\title{On the use of calibration data in error-aware compilation techniques for NISQ  devices}

\author{\IEEEauthorblockN{Handy Kurniawan}
\IEEEauthorblockA{
\textit{Universidad Complutense de Madrid}\\
Madrid, Spain \\
handykur@ucm.es}
\and
\IEEEauthorblockN{Laura Rodríguez-Soriano}
\IEEEauthorblockA{
\textit{Universitat Politècnica de València}\\
Valencia, Spain \\
lrodsor1@disca.upv.es}
\and
\IEEEauthorblockN{Daniele Cuomo}
\IEEEauthorblockA{
\textit{Universitat Politècnica de València}\\
Valencia, Spain \\
cuomo.daniele@outlook.com}
\and
\IEEEauthorblockN{Carmen G. Almudever}
\IEEEauthorblockA{
\textit{Universitat Politècnica de València}
Valencia, Spain \\
cargara2@disca.upv.es}
\and
\IEEEauthorblockN{Francisco García Herrero}
\IEEEauthorblockA{
\textit{Universidad Complutense de Madrid}
Madrid, Spain \\
francg18@ucm.es}
}

\maketitle

\begin{abstract}
Reliably executing quantum algorithms on noisy intermediate-scale quantum (NISQ) devices is challenging, as they are severely constrained and prone to errors. Efficient quantum circuit compilation techniques are therefore crucial for overcoming their limitations and dealing with their high error rates. These techniques consider the quantum hardware restrictions, such as the limited qubit connectivity, and perform some transformations to the original circuit that can be executed on a given quantum processor. Certain compilation methods use error information based on calibration data to further improve the success probability or the fidelity of the circuit to be run. However, it is uncertain to what extent incorporating calibration information in the compilation process can enhance the circuit performance. 
For instance, considering the most recent error data provided by vendors after calibrating the processor might not be functional enough as quantum systems are subject to drift, making the latest calibration data obsolete within minutes. 

In this paper, we explore how different usage of calibration data impacts the circuit fidelity, by using several compilation techniques and quantum processors (IBM Perth and Brisbane). To this aim, we implemented a framework that incorporates some of the state-of-the-art noise-aware and non-noise-aware compilation techniques and allows the user to perform fair comparisons under similar processor conditions. Our experiments yield valuable insights into the effects of noise-aware methodologies and the employment of calibration data. The main finding is that pre-processing historical calibration data can improve fidelity when real-time calibration data is not available due to factors such as cloud service latency and waiting queues between compilation and execution on the quantum backend.

\end{abstract}

\begin{IEEEkeywords}
Quantum compilation, error-aware circuit mapping, algorithm fidelity 
\end{IEEEkeywords}

\section{Introduction}

Quantum systems are susceptible to many sources of noise that introduce errors and degrade the fidelity of the algorithms that are run~\cite{nielsen_quantum_2010}. These error sources permeate every stage of circuit execution. For example, in IBM superconducting processors~\cite{ibminfo}, qubit initialization is associated with error rates typically in the range of $10^{-3}$ to $10^{-2}$. In addition, quantum gates exhibit error rates ranging from $10^{-4}$ to $10^{-2}$ for single-qubit (1Q) gates and from $10^{-3}$ to $10^{-2}$ for two-qubit (\twoq{}) gates. Furthermore, qubit information leaks over time because of decoherence, and the process of measuring qubits introduces noise, with error rates typically in the range of $10^{-3}$ to $10^{-2}$. Also, note that these error rates are not uniform across the processor, but they vary in space (i.e. variability between qubits and links) and time (i.e. between calibration cycles). Due to this error variability in the chip, the fidelity of the executed algorithm may differ depending on the selection of the physical qubits, the operations employed, and the degree of gate parallelism.

The error-proneness of current quantum processors, together with the limited qubit count, positions quantum computing in the Noisy Intermediate-Scale Quantum (NISQ) era, where quantum error correction (QEC) schemes cannot be applied. One promising approach to mitigate noise in NISQ devices is through the use of error-aware quantum circuit compilation techniques. In literature, we can find extensive research on integrating error data from quantum processor's calibration into the compilation process~\cite{murali_noise-adaptive_2019, nation_suppressing_2023, niu_hardware-aware_2020, murali_full-stack_2019, bhattacharjee_muqut_2019, nishio_extracting_2020, liu_not_2022, sharma_noise-aware_2023, tannu_not_2019, wagner_optimized_2024}. Some works focus on finding low error rates by using calibration data only for the initial qubit mapping of qubits~\cite{murali_noise-adaptive_2019, nation_suppressing_2023}, whereas others consider error information for the initial qubit mapping as well as for the qubit routing~\cite{niu_hardware-aware_2020, murali_full-stack_2019, bhattacharjee_muqut_2019, nishio_extracting_2020, liu_not_2022, sharma_noise-aware_2023, tannu_not_2019, wagner_optimized_2024}. Note that both processes may significantly contribute to diminishing errors during compilation by selecting the most connected and most reliable qubits and the shortest paths with the lowest error rates. 


However, it is uncertain to what extent incorporating calibration information in the qubit allocation and routing steps can enhance the circuit performance due to the error variability across the processor. More precisely, the potential improvement will depend on how calibration data is processed and included in the circuit compilation. Most of the infrastructures employed to access the quantum devices are usually based on execution queues within cloud services, which introduce non-negligible delays in the execution of the quantum circuits. For instance, in the case of IBM devices, the queue time \textemdash time elapsed since the circuit is sent until the execution starts\textemdash{} can be up to eight hours~\cite{ibm_fairshare}. For this reason, the error information provided by the vendor may be outdated within minutes after calibration due to the quantum system drifting~\cite{wilson_just--time_2020}. 

In this paper, we explore different methods to process the error information provided in several calibration cycles (historical data) to incorporate it in state-of-the-art noise-aware compilation techniques. The aim is to improve the overall fidelity of the executed circuits even when no real-time error data is available. To conduct the experiments, we developed a framework that allows the user to compare the performance of the subject compilation methods under similar conditions in real backends. The experiments were implemented on two IBM processors (\perth{} and \brisbane{}), showing that using historical calibration data is more beneficial than using the latest available calibration data whenever this is not provided in real time. Specifically, it was observed a circuit fidelity enhancement that increases with the size of the device. For instance, on \perth{} there was an average improvement of 0.08\% (up to 4.09\%) and 8.32\% (up to 41.76\%) on \brisbane{}.



This paper is structured as follows. In Sec.~\ref{sec:background}, we provide the background on error-aware compilation techniques and device calibration information. The methodology used is detailed in Sec.~\ref{sec:methodology} along with the designed framework to execute the noise-aware compilation techniques under fair conditions for comparison. In Sec.~\ref{sec:results}, we derive some observations for the results of the conducted experiments. We conclude with Sec.~\ref{sec:conclusion}, summarizing the main findings and outlining future directions.

\section{Background} \label{sec:background}




\subsection{Quantum circuit compilation process} \label{ssec:compilation_process}

\begin{figure}[htbp]
    \centering
    \begin{subfigure}[b]{.4\linewidth}
        \includegraphics[width=1\linewidth]{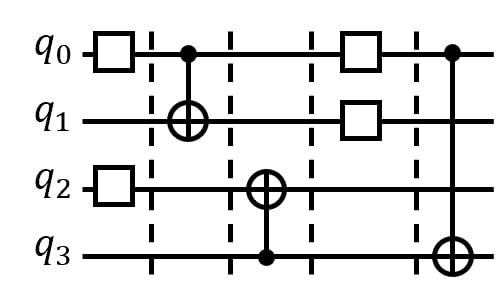}
        \caption{}
        \label{subfig:ex-initial-circuit}
    \end{subfigure}
    \hspace{15px}
    \begin{subfigure}[b]{.35\linewidth}
        \includegraphics[width=1\linewidth]{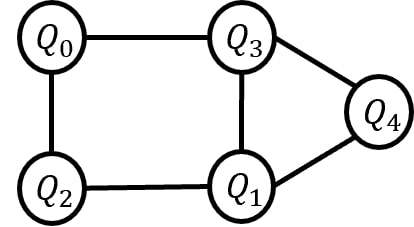}
        \caption{}
        \label{subfig:ex-topology}
    \end{subfigure}

    \begin{subfigure}[b]{.70\linewidth}
        \includegraphics[width=1\linewidth]{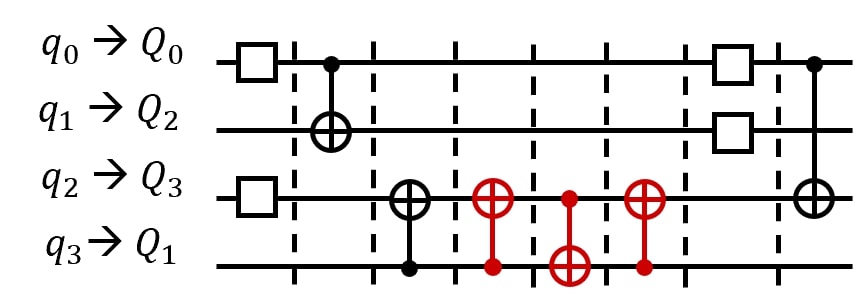}
        \caption{}
        \label{subfig:ex-routing}
    \end{subfigure}
    \caption{Quantum circuit compilation example: (a) original circuit, (b) backend topology, and (c) initial qubit mapping and routing result with the added SWAP gate (in red).}
    \vspace{-1.5em}
    \label{fig:mapping-process}
\end{figure}

Quantum circuit compilation transforms a quantum circuit into an equivalent one that can be executed on a quantum device accordingly with hardware limitations. Fig.~\ref{fig:mapping-process} illustrates some steps of the compilation process with an example: let us consider the circuit depicted in Fig.~\ref{subfig:ex-initial-circuit} (referred to as the original circuit) intended for execution on a device with the topology represented in Fig.~\ref{subfig:ex-topology}; where physical qubits are represented by circles and denoted with $Q_n$ ($n$ corresponds to the qubit index), and the edges represent links to perform \twoq{} gates. The initial qubit placement, also known as qubit allocation or assignment, involves mapping each virtual qubit in the circuit ($q_n$) to a physical qubit on the device ($q_n \xrightarrow{} Q_n$), as illustrated in Fig.~\ref{subfig:ex-routing}. This allocation enables the execution of the first two CNOT gates as there is an edge connecting the respective qubits ($Q_0$ and $Q_2$, $Q_1$ and $Q_3$). However, to execute the third CNOT gate (between $q_0$ and $q_3$), the information of the involved qubits must be moved to adjacent ones. To do this, it is necessary to add SWAP\footnote{\textcolor{black}{For two qubits A and B, SWAP(A,B) := \{CNOT A,B; CNOT B,A; CNOT A,B;\}.}} gates. This is shown in Fig.~\ref{subfig:ex-routing}, where a SWAP between $q_2$ and $q_3$ allows the execution of the third CNOT gate in the initial circuit. 

Other compilation steps not illustrated in the example are i) operation scheduling, to optimize the computation according to a specific metric (e.g. circuit depth), and ii) gate nativization, which decomposes the gates into the native set of gates the device supports. 

\subsection{Prior work} \label{ssec:prior_works}

As can be seen from Fig.~\ref{fig:mapping-process}, the compilation steps usually result in an increase in the number of gates and circuit depth, which generally affects negatively the overall circuit fidelity. Therefore, it is important to use efficient initial qubit mapping and routing strategies that add the fewest SWAP gates; but this is an NP-hard problem~\cite{siraichi_qubit_2018}. For this reason, most of the proposed (and scalable) compilation techniques use heuristics for initial qubit mapping and/or routing~\cite{li_tackling_2019, liu_not_2022, park2022fast, murali_full-stack_2019, murali_noise-adaptive_2019, mckinney_mirage_2023}. For instance, with the reverse traversal method employed in~\cite{li_tackling_2019, liu_not_2022}, the circuit is first executed with a random initial placement strategy. Then the final mapping from this execution is chosen as the initial qubit mapping to run the reversed circuit, and the final qubit mapping is used as the updated initial qubit mapping. An alternative approach for the initial qubit allocation involves using recursive graph-isomorphism search~\cite{nation_suppressing_2023, park2022fast}. This method searches for a subgraph in the topology that is isomorphic to the qubit interaction graph. Regarding the routing of qubits,~\cite{li_tackling_2019} and~\cite{liu_not_2022}, in addition to looking for the shortest path, they use a cost function with a look-ahead window to consider future gates instead of considering the entire circuit. This reduces the search space and allows to make informed decisions and add operations that have minimal impact on subsequent gates. The work in \cite{park2022fast}~analyzes the possibility of using bridge gates (which consist of four operations without swapping qubit information) instead of swap gates to lower gate overhead and consequently reduce noise.

Other quantum circuit compilation techniques have also been proposed to further enhance the reliability of quantum algorithms by taking into account the error characteristics of the quantum device. Some methods consider the measurement error of each qubit to perform the initial qubit mapping when virtual qubits can be placed in different areas of the chip that have the same connectivity, as described in~\cite{murali_noise-adaptive_2019, murali_full-stack_2019}. Additionally, during qubit routing, the fidelity of each qubit coupling link (i.e. two-qubit gate fidelity) is taken into account, ensuring that quantum information is moved through the most reliable links~\cite{murali_full-stack_2019}.

Some vendors provide the necessary error information to apply these compilation techniques after calibrating the quantum devices. However, as mentioned in Section I, this information is outdated shortly after calibration takes place due to the drift that processors suffer from. Obtaining real-time calibration data poses a challenge. The vendor \textcolor{black}{provides} data measured several times a day, e.g., twice a day according to~\cite{yeter-aydeniz_measuring_2023}, and due to the existence of an execution queue, accurately predicting the time between compilation and the moment in which the algorithm will be executed on the real backend remains a limitation, highlighting how hard it is to align experimental conditions with real-time quantum processor operations. \textcolor{black}{An alternative approach was proposed in~\cite{wilson_just--time_2020} to obtain} the latest calibration data through a micro-benchmark and use it to compile the circuit. However, in~\cite{das_imitation_2023}, it is argued that micro-benchmarks are sub-optimal for quantum processors with more than one way of implementing \twoq{} gates. In contrast,~\cite{liu_enabling_2023} presents a real-time framework to re-calibrate unused qubits while running a quantum circuit on the processors. \textcolor{black}{But it is important to note that the current open-source cloud quantum machines do not support flexible and user-defined calibration, which limits the framework's usage.}

\subsection{Calibration data} \label{ssec:calibration_data}

\begin{figure}[htbp]
    \centering
    \includegraphics[width=.99\linewidth]{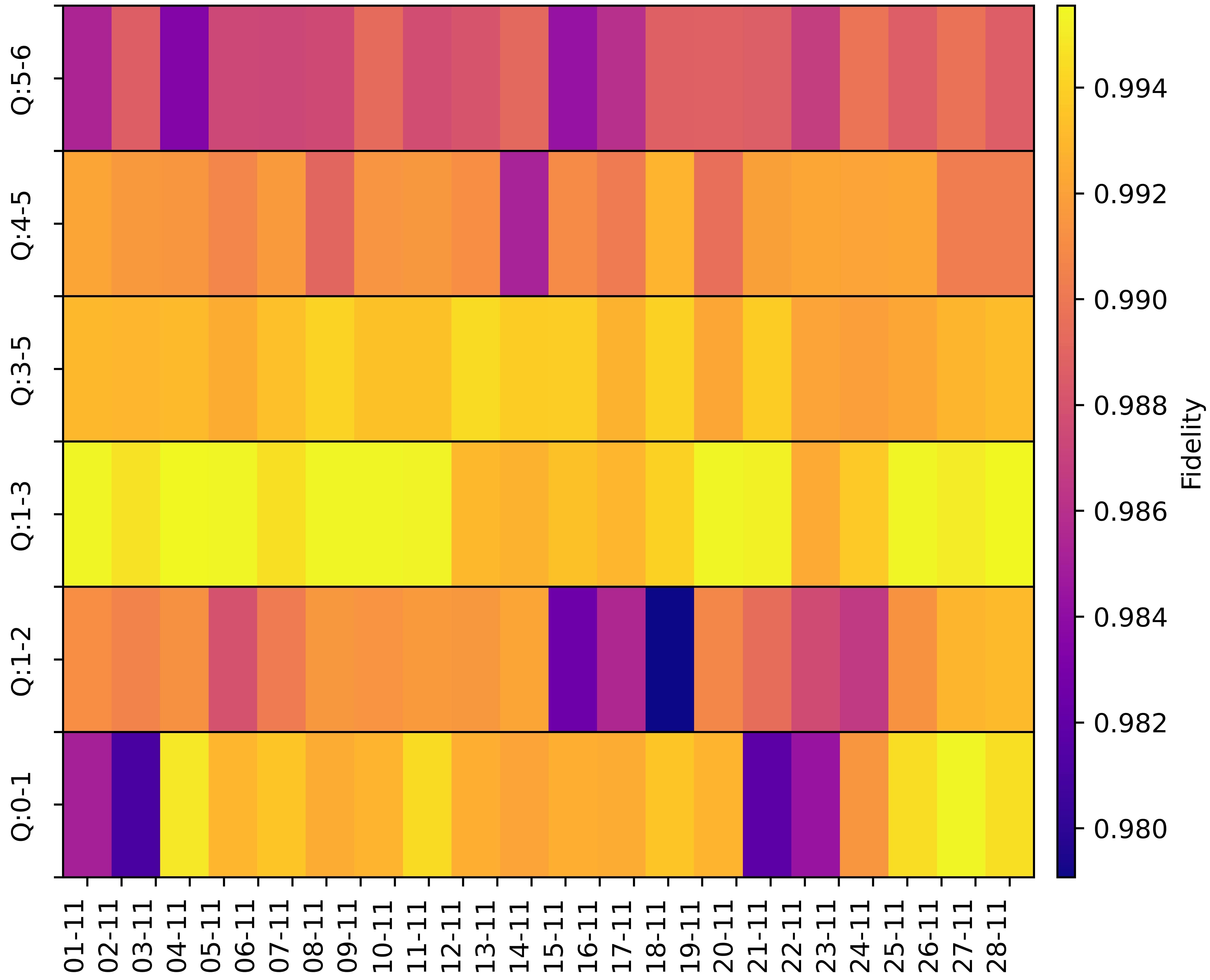}
    \caption{Fidelity of \twoq{} gates between each pair of qubits in \perth{}. The x-axis represents the temporal variability over 28 days, while the y-axis represents the spatial variability of the 6 pairs of qubits.}
    \vspace{-1.5em}
\label{fig:perth_fidelity_variability}
\end{figure}

Noise-aware compilation techniques consider error rate information coming from calibration routines, which provide an error map of the quantum chip at a specific time. This serves as a comprehensive static snapshot of the quantum computer's state. However, the calibration data varies across qubits and in time, meaning there is a spatiotemporal variability in the fidelity of different operations such as 1Q and \twoq{} gates or measurements~\cite{murali_noise-adaptive_2019}. Fig.~\ref{fig:perth_fidelity_variability} displays an example of the reliability of performing a \twoq{} gate between any pair of qubits on the \perth{} chip (see Fig.~\ref{fig:backend_ibm_perth}). The graph illustrates both the spatial (y-axis) and temporal (x-axis, 28 days) variability of the calibration data. For instance,  the fidelity of the \twoq{} gate between qubits 1 and 3 remains consistently high over time, while the \twoq{} gate between qubits 0 and 1 shows greater temporal variability, with some days exhibiting low fidelity (purple) and others showing a significant increase (yellow). Considering the spatial variability, it can be observed that a \twoq{} gate between qubits 5 and 6 generally exhibits lower fidelity than a \twoq{} gate between qubits 1 and 3. Aware of this variability, we investigate how to process and integrate calibration data within error-aware compilation techniques. \textcolor{black}{To this end, error information has been gathered every time there was a new calibration cycle  for two different IBM quantum processors.}

\section{Methodology} \label{sec:methodology}

We conduct experiments on various IBM devices using several compilation techniques, including noise-aware and non-noise-aware methods. We consider different approaches for processing calibration data we collected for several months, with daily acquisitions. As a result, each original circuit is compiled using different qubit mapping and routing strategies. These form a Qiskit job\footnote{A Qiskit job is formed by a collection of circuits that will be executed on the same backend when the user's turn comes up after waiting in the execution queue.} and submitted to the execution queue of the given backend. This protocol helps to work under similar conditions, as circuits of the same job are run consecutively. Finally, different state-of-the-art circuit reliability metrics are computed based on the obtained device measurements.

\subsection{\textcolor{black}{Designed simulation and execution framework}}

\begin{figure}[htbp]
\centerline{
\includegraphics[width=.96\linewidth]{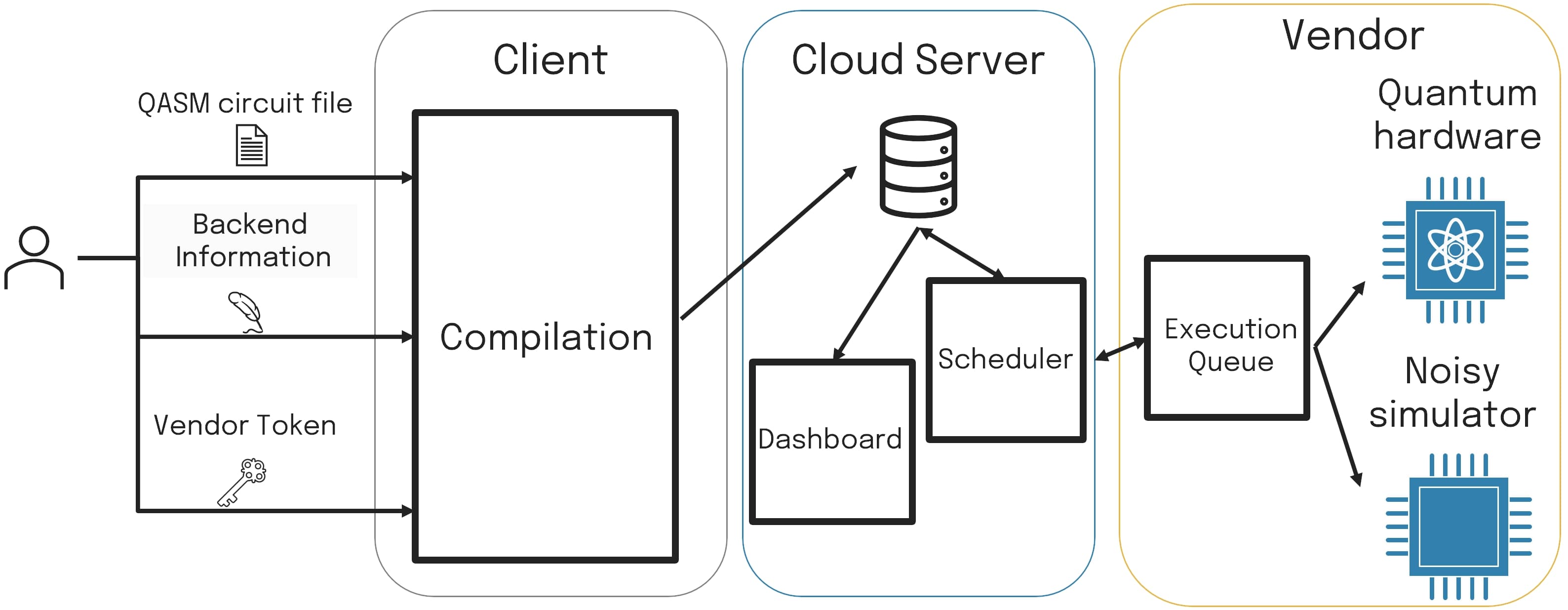}}
\caption{High-level diagram of the designed simulation and execution framework.}
\vspace{-1.5em}
\label{fig:high_level_diagram}
\end{figure}

To conduct the experiments, we developed a framework that aims to compare, under similar conditions, the existing noise-aware compilation methods described in Sec.~\ref{ssec:compilation-techniques}. Previous research, conducted at different times and conditions, lacks cohesive benchmarks and metrics, making it difficult to perform fair and effective performance assessments of compilation. The framework seeks to unify comparisons by providing a consistent tool to evaluate various techniques under the same conditions. It consists of three main components, as shown in Fig. ~\ref{fig:high_level_diagram}: (i) the client application, (ii) the cloud server, and (iii) the vendor's backend. \textcolor{black}{The user inputs an algorithm into the client, selects the quantum processor or noisy simulator, and chooses the compilation strategy to compile the algorithm from the corresponding QASM files. The cloud server submits a quantum job, consisting of all quantum circuits, to the chosen vendor's backend waiting queue.} The results from execution are then processed and used to calculate the metrics defined in Sec.~\ref{ssec:metrics}. Note that in this work, we focused on IBM's quantum hardware (see Sec.~\ref{ssec:device_information}), but the methodology and outcomes may be ported to other vendors and/or compilation techniques and even to quantum error mitigation techniques~\cite{cai_quantum_2022, temme2017error, Nation2021-yl}.

\subsection{Quantum circuit compilation techniques} \label{ssec:compilation-techniques}

Several quantum algorithm compilation techniques can be found in the scientific literature and industry (see Sec.~\ref{ssec:prior_works}). In this study, we focus on two noise-aware techniques: (i) TriQ~\cite{murali_full-stack_2019}; and (ii) Noise-Adaptive (Q-NA)~\cite{murali_noise-adaptive_2019}, implemented in Qiskit. TriQ considers errors in both the initial qubit mapping and routing stages. Noise-Adaptive, on the other hand, only considers them in the initial qubit mapping stage. These are state-of-the-art noise-aware compilation techniques. Note that other methods like NASSC~\cite{liu_not_2022} faced some problems like incompatibility with newer versions of Qiskit, and therefore it was not included in this work. 

To further explore noise-aware routing algorithms, we upgraded TriQ -- named TriQ+ -- by changing the path-finding method. More precisely, instead of using Dijkstra's~\cite{dijkstra_note_1959}, which seeks the shortest path from only one side (moving source to target qubit or vice versa), we use Floyd-Warshall's~\cite{floyd_algorithm_1962}, which considers paths from both source and target at the same time, finding the position of qubits within the shortest path that results in the highest reliability.


The noise-aware approaches were also compared against non-noise-aware compilation techniques. Specifically, Qiskit with optimization level 3 combined with SABRE~\cite{li_tackling_2019} (Q-3)~\cite{qiskit_contributors_qiskit_2023}. At the time of writing, this is the default setting in Qiskit and the one that considers a bigger search space when making decisions about qubit routing. Note that there are other non-noise-aware compilation strategies, such as MIRAGE~\cite{mckinney_mirage_2023}, that we opted to not include in the paper due to obtaining lower fidelity than SABRE. A summary of the aforementioned quantum circuit compilation techniques is shown in Table~\ref{tab:technique_classification}. 

\begin{table}[htbp]
\caption{Features of the quantum circuit compilation techniques considered in this work. }
\begin{center}
\begin{tabular}{|c|c|c|c|c|}
\hline
\textbf{Technique}&\multicolumn{2}{|c|}{\textbf{Initial Qubit Mapping}} &\multicolumn{2}{|c|}{\textbf{Routing}}\\
\cline{2-5} 
 & \textbf{Method}& \textbf{NA}& \textbf{Method} & \textbf{NA} \\ \hline

\multicolumn{1}{|l|}{TriQ~\cite{murali_full-stack_2019}}   & \multicolumn{1}{l|}{Reliability Matrix} & Yes & \multicolumn{1}{l|}{Dijkstra} & Yes \\ \hline

\multicolumn{1}{|l|}{TriQ+~\cite{murali_full-stack_2019}}   & \multicolumn{1}{l|}{Reliability Matrix} & Yes & \multicolumn{1}{l|}{Floyd-Warshall} & Yes \\ \hline

\multicolumn{1}{|l|}{Q-NA~\cite{murali_noise-adaptive_2019}} & \multicolumn{1}{l|}{Noise-adaptive} & Yes & \multicolumn{1}{l|}{SABRE} & No \\ \hline

\multicolumn{1}{|l|}{NASSC~\cite{liu_not_2022}} & \multicolumn{1}{l|}{SABRE} & No & \multicolumn{1}{l|}{NASSC} & Yes  \\ \hline

\multicolumn{1}{|l|}{Q-3~\cite{qiskit_contributors_qiskit_2023}} & \multicolumn{1}{l|}{SABRE} & No & \multicolumn{1}{l|}{SABRE} & No \\ \hline

\multicolumn{1}{|l|}{MIRAGE~\cite{mckinney_mirage_2023}} & \multicolumn{1}{l|}{SABRE} & No & \multicolumn{1}{l|}{MIRAGE} & No \\ \hline

\hline
\multicolumn{5}{l}{${\mathrm{NA}}$: technique is noise-aware.}

\end{tabular}
\label{tab:technique_classification}
\end{center}
\vspace{-1.5em}
\end{table}

All experiments were conducted using Python 3.11.5 within a Docker system equipped with 13 GB of main memory, 8 CPUs, and Ubuntu 20.04. It follows the list of software specifics we used: Qiskit versions 0.44.3~\cite{qiskit_contributors_qiskit_2023}, qiskit-aer versions 0.12.3~\cite{github_qiskit-aer}, qiskit-ibm-runtime versions 0.12.2~\cite{github_qiskit-ibm-runtime}, Mirage versions 0.2.0~\cite{github_mirage}, and updated TriQ package~\cite{github_triq}.

\subsection{\textcolor{black}{Historical} calibration data} \label{ssec:calibration_data_sets}


Motivated by the fact that the latest calibration data is often outdated at the time of executing a job (i.e., the waiting time can be up to eight hours~\cite{ibm_fairshare}), we explore alternatives to exploit the error information available. Specifically, we collected and analyzed calibration data from IBM backends, searching for a methodology to better exploit all this information to enhance the fidelity of circuits when real-time calibration data is unavailable.

Our proposal is to process historical calibration data and apply noise-aware compilation techniques accordingly. The data collection spans over several months, with a daily basis extraction. We classify the error information as follows: i) \textit{Latest (lcd)} that considers the calibration data that is provided by the vendor at the moment of sending the quantum circuit to the backend waiting queue; ii) \textit{Average (avg)} in which the mean of all the stored error rates is used; iii) \textit{Mix (mix)} that combines the last calibration data with the average for those qubits that have not been calibrated the same day that the experiment is sent to the backend waiting queue; and iv) \textit{Window-$n$ (w-$n$)}  calculates the average fidelity of a window size of $n$ days prior to the day the circuit is added to the queue. On top of this, taking into account temporal variability in the calibration data, we calculate the \textit{standard deviation (std)} of the collected data and apply it to the above methods (listed in Table~\ref{tab:calibration_set_classification}). This processing method is referred to as \textit{-adj} in the result section. For instance, \textit{lcd-adj} means the latest calibration data considering the standard deviation. \textcolor{black}{This involves generating a new metric which is a linear combination of the error rate and the standard deviation derived from historical calibration data for each pair of gates and qubits} High deviation can make the qubit less reliable, with a fluctuating error rate. Conversely, low deviation can provide a more stable and reliable qubit. The goal is to penalize qubit pairs with high deviation and select the more reliable one.

\begin{table}[htbp]
\caption{List of methods to process calibration data considered in this paper. }
\begin{center}
\begin{tabular}{|c|c|}
\hline
\textbf{Name}&\textbf{Description} \\ \hline

lcd &  \multicolumn{1}{l|}{Latest calibration data available at the vendors database} \\ \hline
avg & \multicolumn{1}{l|}{Average fidelity from the beginning of the experiments.} \\  
 & \multicolumn{1}{l|}{to the day that the circuit is sent to the backend waiting queue} \\ \hline
mix &
  \multicolumn{1}{l|}{If the qubit has not been calibrated today, utilize} \\
  & \multicolumn{1}{l|}{the avg fidelity.  Otherwise, use the latest calibration data.} \\ \hline
w-$n$ & \multicolumn{1}{l|}{Average fidelity with a window size of $n$ days.}  \\ \hline
\end{tabular}
\label{tab:calibration_set_classification}
\end{center}
\vspace{-1.5em}
\end{table}

\subsection{Benchmarks} \label{ssec:benchmarks}


To assess the accuracy of our proposed calibration processing methods, we selected a set of quantum circuits with a high percentage of \twoq{} gates, as summarized in Table~\ref{tab:benchmark_list}. It is important to note that these circuits have been used as benchmarks in previous studies~\cite{murali_full-stack_2019, li_qasmbench_2022, bandic_interaction_2023}. This study aims to analyze the impact of noise-aware compilation techniques on the overall performance of quantum algorithms under standardized conditions. 

\begin{table}[htbp]
\caption{Summary of benchmarks used in our study.}
\begin{center}
\begin{tabular}{|c|c|c|c|c|c|}
\hline
\textbf{Circuit}&\textbf{Qubits}&\textbf{Gates}&\textbf{2Q Gates}&\textbf{2Q Gates(\%)}&\textbf{Depth} \\ \hline

Adder & 4 & 34 & 14 & 41.18 & 24\\ \hline
And & 5 & 109 & 36 & 33.03 & 83\\ \hline
BV* & 2 & 6 & 1 & 16.67 & 5\\ \hline
BV* & 3 & 9 & 2 & 22.22 & 6\\ \hline
BV & 4 & 12 & 3 & 25 & 7\\ \hline
BV & 5 & 15 & 4 & 26.67 & 8\\ \hline
BV & 6 & 18 & 5 & 27.78 & 9\\ \hline
BV & 7 & 21 & 6 & 28.57 & 10\\ \hline
BV & 8 & 24 & 7 & 29.17 & 11\\ \hline
BV & 9 & 27 & 8 & 29.63 & 12\\ \hline
BV & 10 & 30 & 9 & 30 & 13\\ \hline
BV & 11 & 33 & 10 & 30.3 & 14\\ \hline
BV & 12 & 36 & 11 & 30.56 & 15\\ \hline
Fredkin & 3 & 19 & 8 & 42.11 & 12\\ \hline
HS4 & 4 & 28 & 4 & 14.29 & 10\\ \hline
Or & 5 & 116 & 36 & 31.03 & 83\\ \hline
QFT & 5 & 61 & 26 & 42.62 & 35\\ \hline
Toffoli & 3 & 19 & 6 & 31.58 & 13\\ \hline
\multicolumn{6}{l}{$^{\mathrm{*}}$ The circuit doesn't need routing.} \\
\end{tabular}
\label{tab:benchmark_list}
\end{center}
\vspace{-2.5em}
\end{table}

\subsection{Quantum processors}\label{ssec:device_information}

\begin{figure}[htbp]
\centering
\begin{subfigure}[b]{0.24\textwidth}
  \centering
  \includegraphics[scale=0.29]{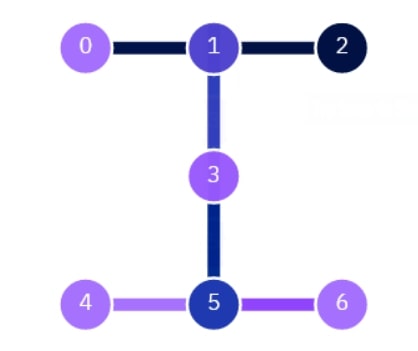}
  \caption{}
  \label{fig:backend_ibm_perth}
\end{subfigure}%
\hfill
\begin{subfigure}[b]{0.24\textwidth}
  \centering
  \includegraphics[scale=0.13]{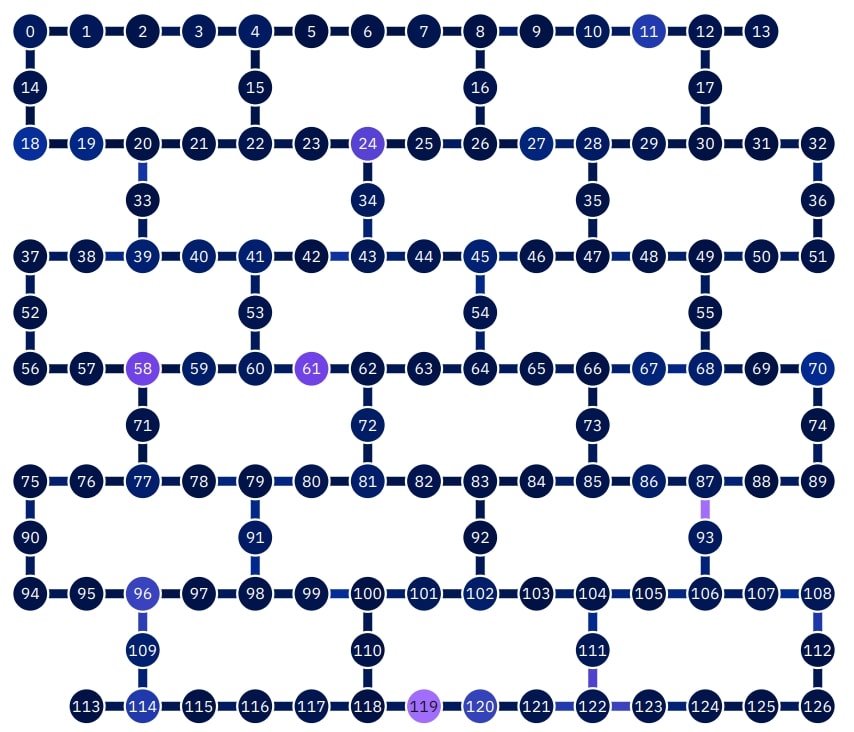}
  \caption{}
  \label{fig:backend_ibm_brisbane}
\end{subfigure}
\caption{IBM quantum processors topology: (a) \perth{}, and (b) \brisbane{}~\cite{ibminfo}}
\vspace{-1.5em}
\label{fig:backend_ibm}
\end{figure}

IBM has several quantum chips based on superconducting technology that differ in qubit size, topology, set of native gates, and processor architecture. In this work, we focused on two (openly available) IBM devices that exemplify this diversity. Firstly, we used \textit{\perth{}} (Fig. ~\ref{fig:backend_ibm_perth}), which has 7 qubits and is powered by the Falcon r5.11 processor. Its supported \twoq{} native gate is the controlled-NOT (CNOT)~\cite{ibm_cxgate}. Secondly, we used \textit{\brisbane{}} (Fig.~\ref{fig:backend_ibm_brisbane}), which has 127 qubits and employs the Eagle r3 processor with Echoed Cross-Resonance (ECR)\footnote{ECR is equivalent to a CNOT up to single-qubit pre-rotations \cite{ibm_ecrgate}.} gate as the \twoq{} native gate. \textcolor{black}{This choice of quantum processors was made based on availability and size. Larger devices offer more options for the initial qubit placement and routing leading to different results. }

After selecting the backend, the circuit was run for 8192 shots\footnote{When we started designing the framework under evaluation, the maximum number of shots was 8192. Now the number of shots has increased, but for consistency, we keep the same number.}. To increase precision, each circuit was replicated four times, and the results were compiled into one job. This job was submitted to the selected backend daily for several days, ranging from 7 to \textcolor{black}{14 days}. Subsequently, from the executed circuits, we then derived different performance metrics as explained in Sec.~\ref{ssec:metrics}.

\begin{table}[ht]
\caption{Methods to calculate fidelity used in this work. $N$ is the number possible outcomes $N = 2^n$, where $n$ is the number of qubits. $P_{ \text{noisy}}$ corresponds to the noisy probability distribution from a real device, and $P_{ \text{noiseless}}$ is the probability distribution from a noiseless simulator. Note that the first metric can only be used to measure the fidelity of deterministic circuits, and the last two metrics can be applied to deterministic and non-deterministic algorithms.}
    \centering
    \begin{tabular}{|@{}c@{}|@{}c@{}|}
    \hline
        \textbf{Metric} & \textbf{Formula} \\
    \hline
        Fidelity~\cite{liu_not_2022} & 
        \centered{\\{ $P_{ \text{noisy}}/P_{ \text{noiseless}}$}\\\\} \\ 
    \hline
        Fidelity~\cite{patel_quest_2022} & 
        \centered{\\{\normalsize $ 1 - \left ( \frac{1}{2} \sum_{k=1}^{N} \left| P_{ \text{noisy}}(k) - P_{ \text{noiseless}}(k) \right| \right )$}\\\\}\\
    \hline
        ~Correlation~\cite{das_case_2019}~~ & 
        \centered{\\{\normalsize ~$1 - \frac{1}{\sqrt{2}} \sqrt{\sum_{k=1}^{N} \left ( P_{ \text{noisy}}(k) - P_{ \text{noiseless}}(k)  \right )^2}$~}\\\\}\\
    \hline
    \end{tabular}
     \label{tab:formulas}
\end{table}

\subsection{Noisy simulator} \label{ssec:noisy_simulator}

To emulate the effect of having access to real-time error data, we used the \texttt{ibmq\_qasm\_simulator}~\cite{ibminfo}, building models that replicate the topology of real backends and load all the calibration data from the device. With this method, we can emulate the behaviour of a quantum device that behaves exactly in the same manner as the real device at the moment that it was calibrated. It is important to highlight that the simulator does not consider a phenomenological noise model with just a single physical error rate parameter, but the exact calibration data of each of the parts that compose the device, including readout circuits, individual quantum gates, etc. 


\subsection{Performance metrics} \label{ssec:metrics}


To evaluate the performance of different quantum circuit compilation techniques, various metrics are typically used, including 1Q and \twoq{} gates overhead, circuit depth, and a fidelity $F$ or Estimated Success Probability $ESP$. In this work, we chose \twoq{} gate overhead, circuit depth, and fidelity as performance metrics. Note that to compute the circuit fidelity, we use three different equations based on ~\cite{patel_quest_2022, liu_not_2022, das_case_2019}, \textcolor{black}{which are shown in Table \ref{tab:formulas}.}

\section{Results} \label{sec:results}

The results presented in this section are based on experiments conducted on two real backends, \perth{}  and \brisbane{} (refer to Sec.~\ref{ssec:device_information}), as well as a noisy simulator (Sec.~\ref{ssec:noisy_simulator}). \textcolor{black}{Each of the techniques described in Sec.~\ref{ssec:compilation-techniques} was evaluated using the circuits listed in Table~\ref{tab:benchmark_list}. For details on the number of runs and shots, refer to Sec.~\ref{ssec:device_information}.}


By performing an exhaustive analysis, we derived several observations that could help to further understand the performance of different combinations of quantum circuits, noise-aware compilation techniques, and quantum devices. We expect that these initial observations, together with the developed framework, can assist in getting a deeper knowledge of the use of calibration information for mitigating quantum errors and enhancing the performance of quantum algorithms when executed on NISQ devices. 

\begin{enumerate}
    \item Noise-aware compilation techniques based on calibration data can enhance the fidelity of the quantum algorithms when calibration data \textcolor{black}{is properly considered.} 
    
    \item In addition, the closer the calibration data is to the real-time data, the larger the improvement of the circuit fidelity when using noise-aware compilation techniques. 

    \item In the absence of real-time calibration data, noise-aware compilation techniques that rely solely on the latest calibration data can result in greater deviation from the theoretical results. In contrast, utilizing historical calibration data can prevent this deviation.
    
    \item Compilation strategies based on calibration data can lead to higher circuit fidelities than those only optimizing the number of added \twoq{} gates. In fact, using more gates (i.e., SWAPs) may still result in a higher circuit fidelity if these are properly selected according to calibration data.

    \item Noise-aware compilation techniques can lead to higher fidelity enhancements when applied to larger devices.

    \item \textcolor{black}{Among the noise-aware techniques, no strategy emerges as superior in enhancing circuit fidelity.} All of them provide improvements whose magnitude depends on the structure of the quantum circuit under study and the targeted quantum device.
\end{enumerate}

Some of these observations were made in previous works. However, they were somehow scattered or even sometimes contradictory, and no clear conclusions could be derived regarding the use of calibration data in different noise-aware compilation methods. Therefore, in our case, we have performed a thorough evaluation in which all experiments have been conducted under the same system conditions and for different devices and circuit compilation techniques. Authors in~\cite{murali_noise-adaptive_2019, bhattacharjee_muqut_2019, hua_caqr_2023} highlighted the importance of using calibration data in noise-aware compilation techniques, especially during the initial qubit mapping and routing stages. This statement aligns with observation (1). However, we found, and it is also supported by~\cite{das_imitation_2023}, that the latest calibration data may not accurately capture device error characteristics due to frequent device drift. On the other hand, it has been observed that the overall circuit fidelity increases as its execution gets closer to the calibration cycle~\cite{wilson_just--time_2020}. This observation is consistent with our findings in (2). Furthermore, studies~\cite{wilson_just--time_2020, liu_enabling_2023, das_imitation_2023} have shown that frequent and rapid machine drift can cause inaccurate calibration data, causing a compiler to follow deceptive information. This observation is consistent with (3), which shows that relying on the latest calibration data without real-time error information is sub-optimal. Moreover,~\cite{tannu_not_2019, sharma_noise-aware_2023} demonstrate that using higher fidelity qubits, even if it results in longer paths, might lead to better circuit fidelity than simply minimizing circuit size and depth. This finding is consistent with (4). Other studies~\cite{murali_full-stack_2019, sharma_noise-aware_2023} show that higher qubit connectivity and qubit count can enhance circuit fidelity through compilation techniques, which aligns with our observation (5). As expected and mentioned in (6), the level of improvement in circuit fidelity using different noise-aware solutions under the same conditions depends on the algorithm characteristics and targeted processor~\cite{bandic_interaction_2023}.  
Lastly, to the best of our knowledge, observation (3)  has not been reported before. We conclude that using historical calibration data in noise-aware compilation techniques results in a higher reliability improvement. 


In the the following subsections, we will answer a series of questions, based on our experiments, which led to the previous summarized observations.


\subsection{\textcolor{black}{Do} noise-aware compilation techniques improve the fidelity of the algorithm to be executed?}

\begin{figure}[t]
\centering
\begin{subfigure}[b]{0.24\textwidth}
  \centering
  \includegraphics[width=.99\linewidth]{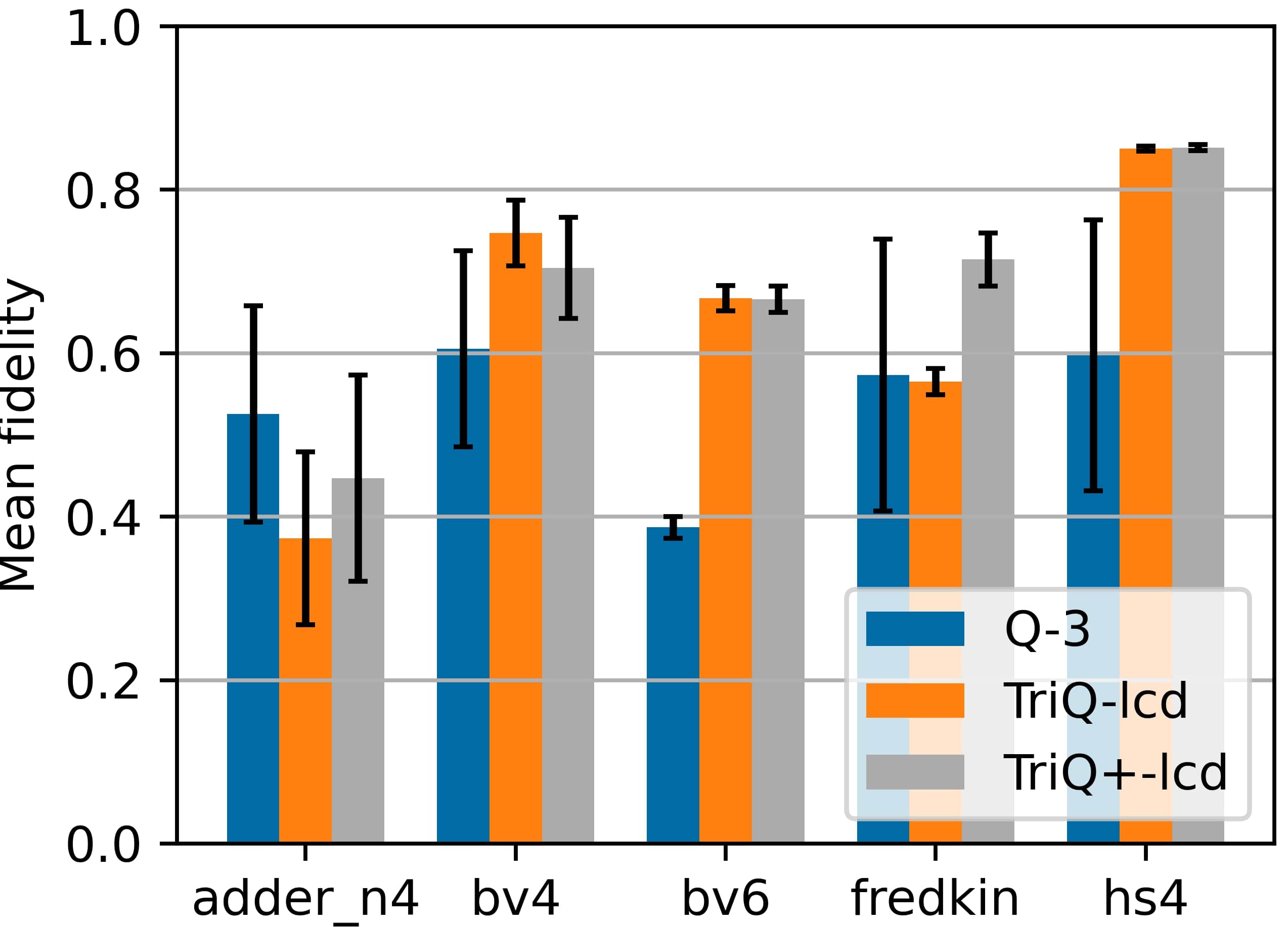}
  \caption{}
  \label{fig:mean_fidelity_compare_noise_aware_with_non_perth}
\end{subfigure}
\begin{subfigure}[b]{0.24\textwidth}
  \centering
  \includegraphics[width=.99\linewidth]{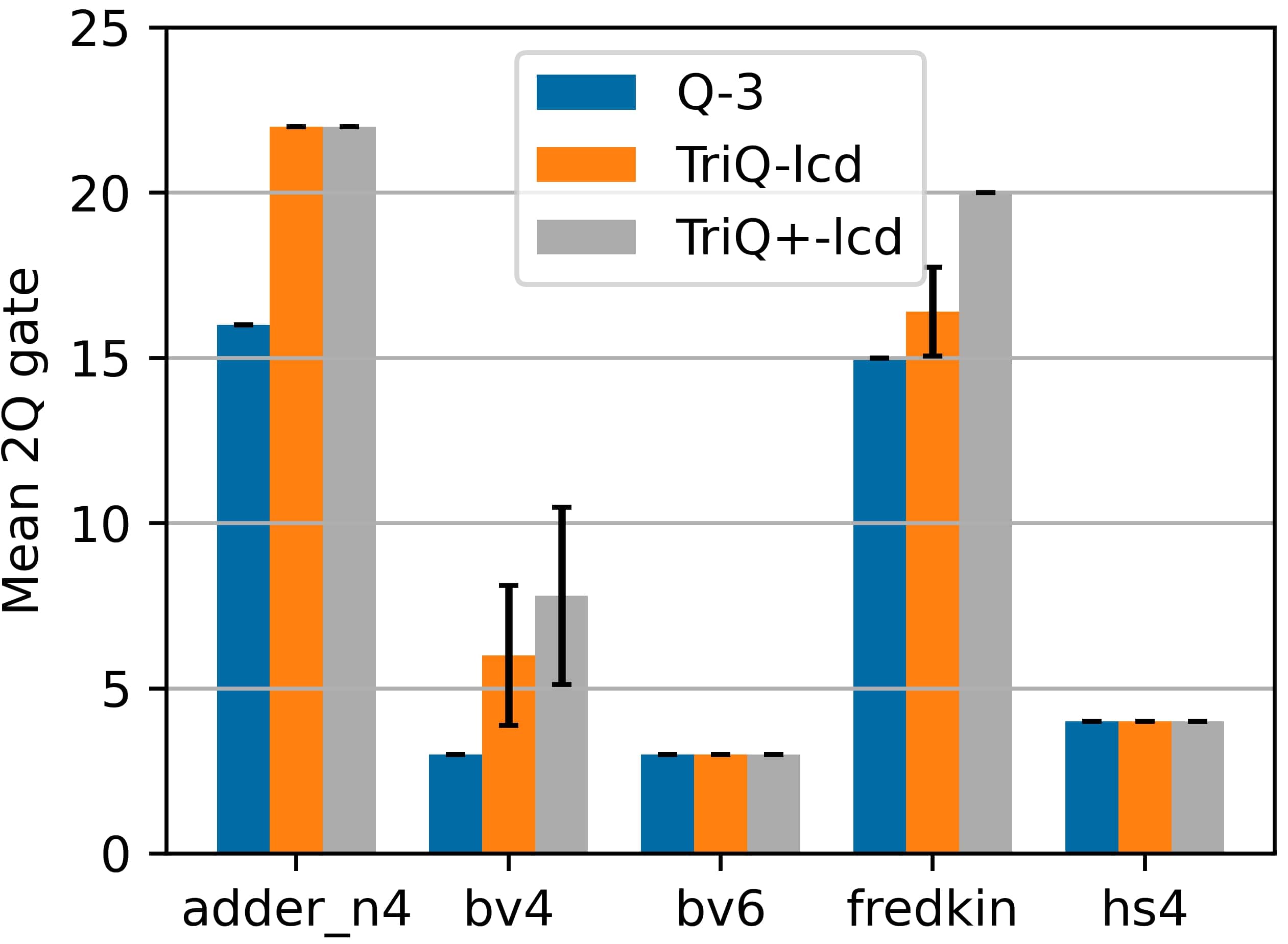}
  \caption{}
  \label{fig:mean_2q_compare_noise_aware_with_non_perth}
\end{subfigure}

\begin{subfigure}[b]{0.24\textwidth}
  \centering
  \includegraphics[width=.99\linewidth]{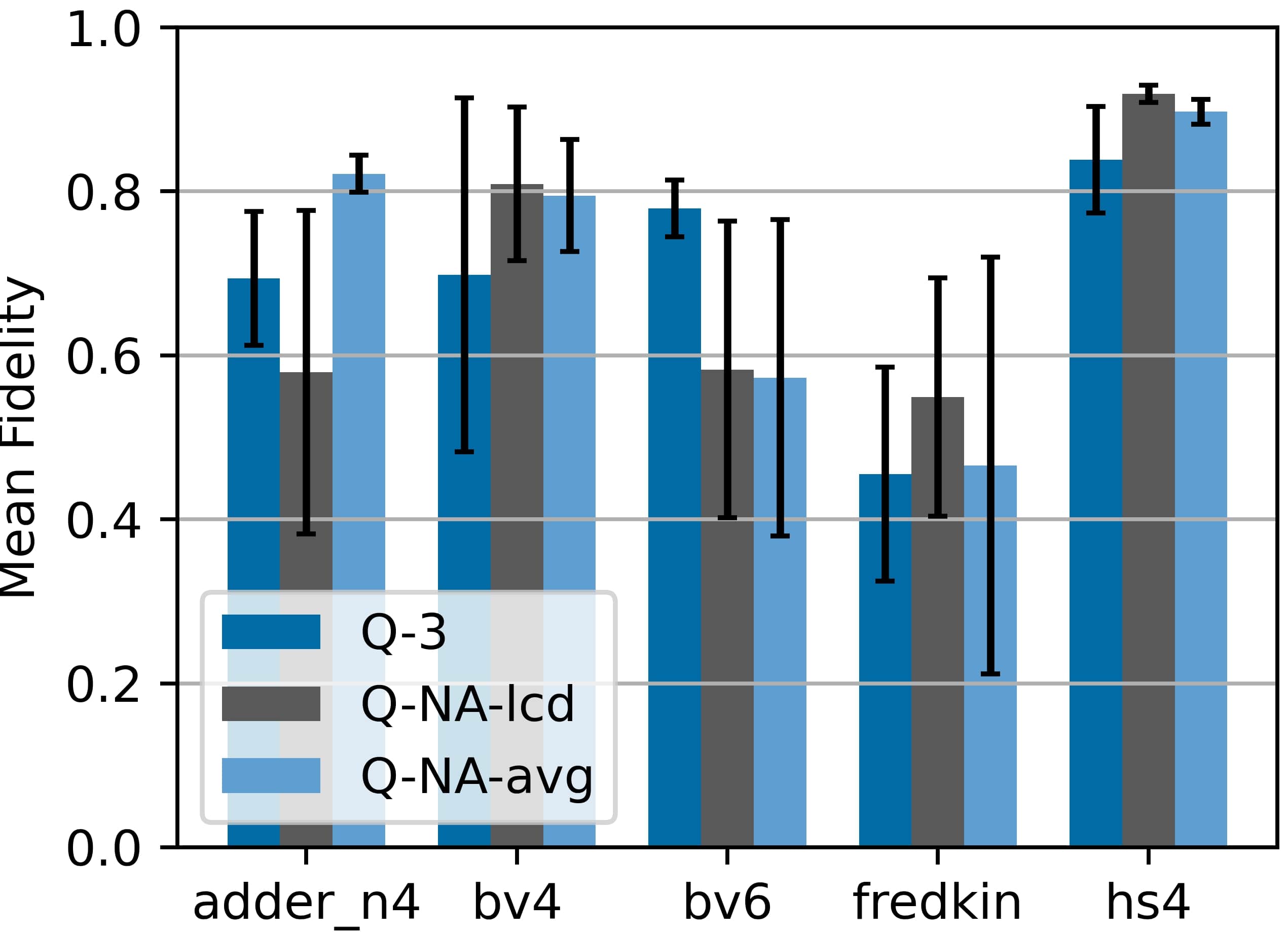}
  \caption{}
  \label{fig:mean_fidelity_compare_noise_aware_with_non_brisbane}
\end{subfigure}
\begin{subfigure}[b]{0.24\textwidth}
  \centering
  \includegraphics[width=.99\linewidth]{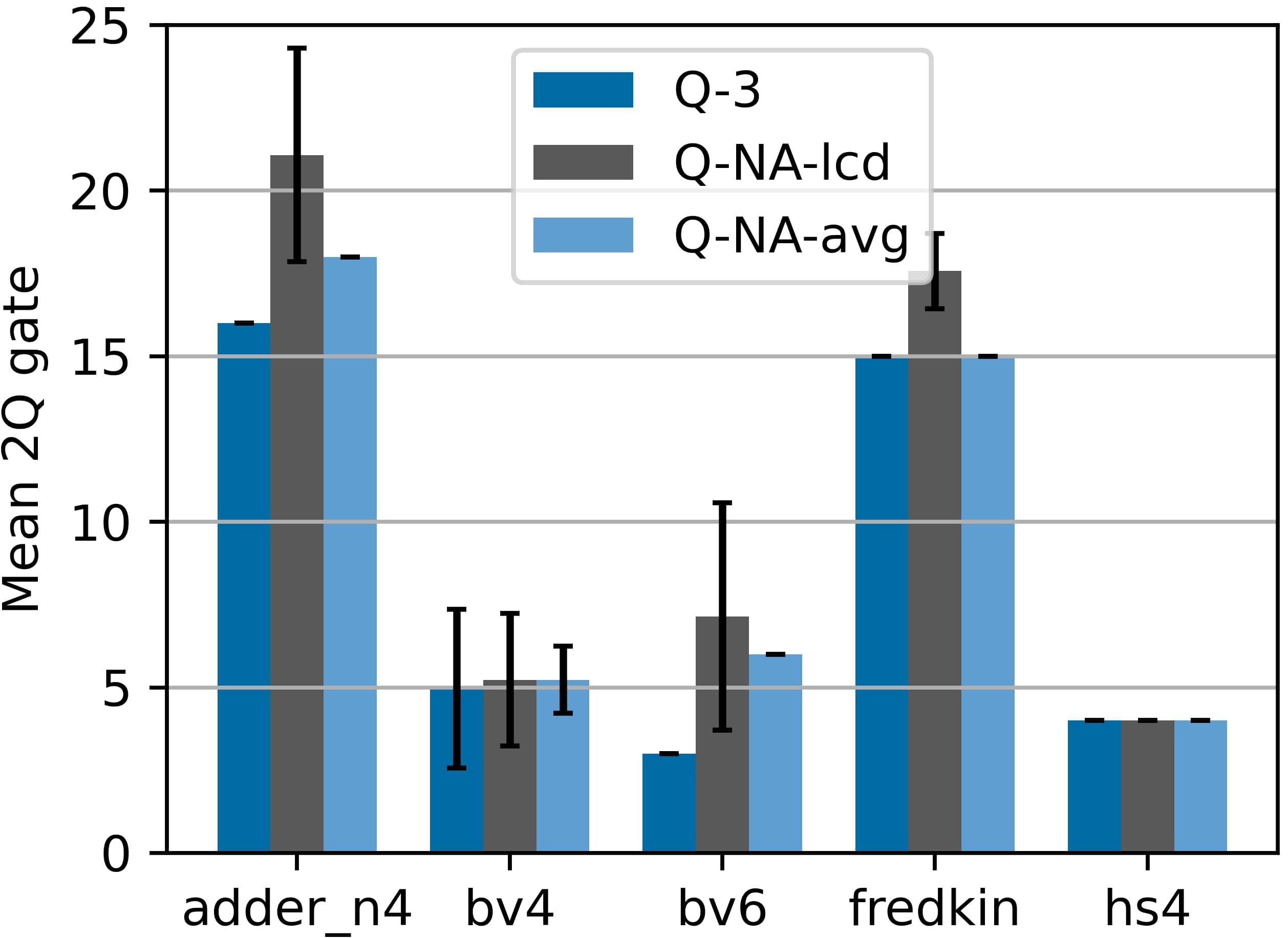}
  \caption{}
  \label{fig:mean_2q_compare_noise_aware_with_non_brisbane}
\end{subfigure}

\caption{\textcolor{black}{Comparison of a non-noise-aware compilation technique (Qiskit with optimization level 3 in dark blue) and noise-aware techniques (\textcolor{black}{orange} for TriQ using last calibration data,  \textcolor{black}{light grey} for TriQ+ with last calibration data, \textcolor{black}{dark grey} for Q-NA using last calibration data, and \textcolor{black}{light blue} for Q-NA with average calibration data): (a) Mean fidelity on \perth{}, (b) Mean 2Q gates on \perth{}, (c) Mean fidelity on \brisbane{}, and (d) Mean 2Q gates on \brisbane{}}.}
\vspace{-1.7em}
\label{fig:compare_noise_aware_with_non_noise_aware}
\end{figure}

\textcolor{black}{To analyze how the use of calibration data impacts the potential improvements of noise-aware compilation techniques, we conducted experiments on both \perth{} and \brisbane{} (see Fig.~\ref{fig:compare_noise_aware_with_non_noise_aware}) using noise-aware TriQ and TriQ+ with lcd and avg methods of processing calibration data as the initial exploratory, and non-noise-aware Q-3 as the baseline. However, for \brisbane{}, TriQ and TriQ+ were excluded as they do not scale in finding the best initial qubit mapping with the Z3 solver~\cite{de2008z3}. As an alternative, we applied Q-NA. The results on \perth{}, shown in Fig.~\ref{fig:mean_fidelity_compare_noise_aware_with_non_perth}, indicate that noise-aware compilation techniques (TriQ, TriQ+) generally outperform the non-noise-aware compilation technique (Q-3). In bv4, noise-aware compilation techniques have an overhead of 100\% in the number of \twoq{} gates, as shown in Fig.~\ref{fig:mean_2q_compare_noise_aware_with_non_perth}. However, the circuit fidelity is still higher \textcolor{black}{by 10\%} compared to the non-noise-aware compilation technique. To achieve higher fidelity, it is important to select the most reliable (according to measurement fidelity) qubits for initial qubit mapping, even if it results in an increase in the number of \twoq{} gates. The Fredkin circuit exhibits similar behaviour, with TriQ+ having the highest number of \twoq{} gates among the three, yet still achieving higher fidelity. 
As shown in Fig.~\ref{fig:mean_2q_compare_noise_aware_with_non_perth}, Q-3 outperforms TriQ and TriQ+ when the number of \twoq{} gates in the compiled circuit is higher than 40\% with respect to the baseline circuit. Finally, Fig.~\ref{fig:mean_fidelity_compare_noise_aware_with_non_brisbane} presents results from \brisbane{} that support the previously mentioned trend. The noise-aware compilation technique (Q-NA) results in a higher fidelity for all circuits except bv6, where the increased number of \twoq{} gates is twice as high compared to Q-3. However, the observation still holds for circuit adder\_n4 and Fredkin gate, where the increase of \twoq{} gates is just around 20\%. The difference between Q-3 and Q-NA is due to the initial qubit mapping process. It is important to note that a higher number of \twoq{} gates can still lead to an increased fidelity, as shown in Fig.~\ref{fig:mean_2q_compare_noise_aware_with_non_brisbane}, with the exception of bv6. The results of the algorithms that do not need routing (bv4, bv6, and hs4) emphasize the significance of implementing noise-aware techniques during the initial qubit mapping process. The experiment concludes that \textbf{the noise-aware compilation technique has the potential to improve overall fidelity across various algorithms and topologies. Additionally, it suggests that the number of \twoq{} gates is important, but it does not completely determine reliability}.}


\subsection{Is including the latest calibration data in the compilation method the most efficient way to improve circuit fidelity when real-time error information is not available? }


The objective is to evaluate how incorporating the latest calibration data provided by the vendor in the noise-aware compilation techniques affects the fidelity of the executed circuits. \textcolor{black}{Reference~\cite{yeter-aydeniz_measuring_2023} analyzed the inter-day and intra-day calibration drift of qubits and \cite{wilson_just--time_2020} demonstrated that a rapid drift occurs within minutes, resulting in a 2\% frequency drift and an average 18\%±5\% increment in single-qubit gate errors~\cite{liu_enabling_2023}.} 

\textcolor{black}{Performing this evaluation is challenging because} the job submitted to the quantum processor will be added to the queue, and its execution time cannot be controlled, resulting in a waiting time of up to eight hours~\cite{ibm_fairshare}. A delay in executing the job may cause outdated calibration data for the circuit compilation process. It is hypothesized that the use of obsolete calibration data will not improve the overall fidelity, while the use of calibration data that is closer to real-time error information will increase the overall fidelity.

\begin{figure*}[t]
\centering
\begin{subfigure}[b]{0.44\linewidth}
  \centering
  \includegraphics[width=1\linewidth]{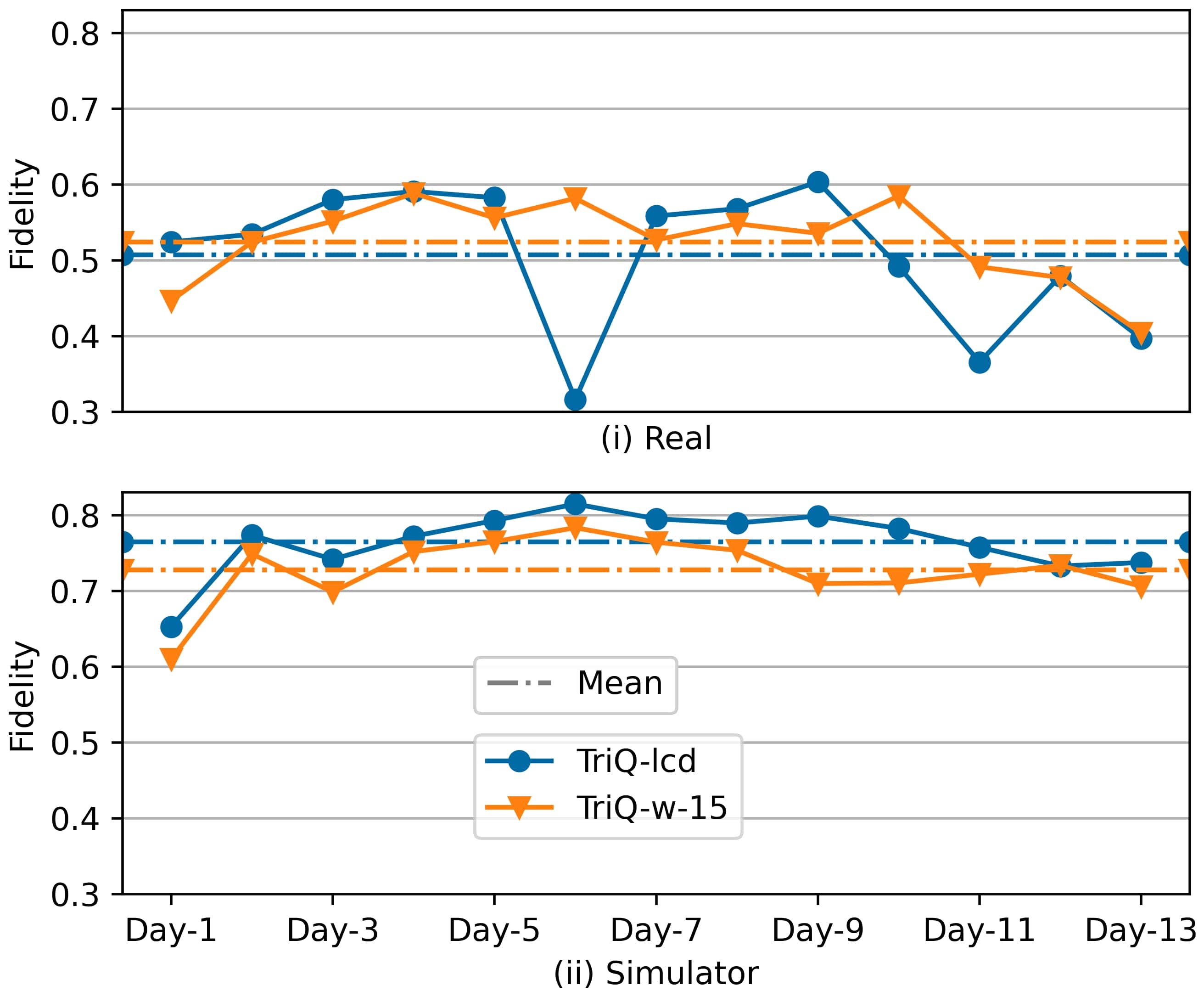}
  \caption{}
  \label{fig:daily_line_perth_adder}
\end{subfigure}
\begin{subfigure}[b]{0.48\linewidth}
  \centering
  \includegraphics[width=1\linewidth]{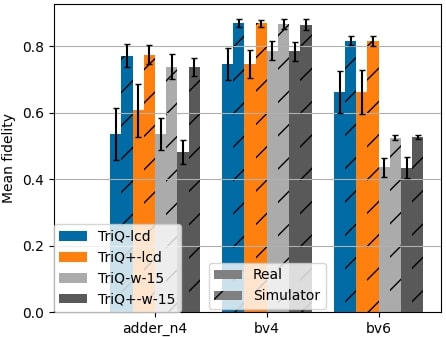}
  \caption{}
  \label{fig:summary_triq_sim_vs_real}
\end{subfigure}
\caption{(a) Daily variability of the adder\_n4 fidelity when executed on the \perth{} quantum processor (top) and simulated in a noisy simulator (bottom) with the noise model extracted from the latest calibration data of \perth{} \textcolor{black}{using the same noise-aware compilation technique (TriQ)}. It emulates having access to real-time error information. (b) Average fidelity for different circuits when run on \perth{} (plain bar) and simulated (dashed bar) using TriQ (\textcolor{black}{blue and light grey}) and TriQ+ (\textcolor{black}{orange and dark grey}) noise-aware compilation techniques.  In this case, both the latest calibration data available at compilation time (\textcolor{black}{blue and orange}) and a window of 15 days of historical calibration data (\textcolor{black}{light and dark grey}) are considered.}
\vspace{-1.5em}

\label{fig:lcd_summary}
\end{figure*}

The top part of \textcolor{black}{Fig.~\ref{fig:daily_line_perth_adder}} (Real) shows that, for a real backend, the fidelity of the quantum circuit under test (adder\_n4) drops more than 30\% (day 6) when only the latest calibration data is considered in the error-aware compilation technique (TriQ-lcd). On the bottom \textcolor{black}{(simulator)}, we can observe that the noise-aware compilation technique exhibits a better and more stable performance with the latest calibration data when simulating the quantum circuit. This is because the noisy simulator uses exactly the same (latest) calibration data provided to the compiler, which \textcolor{black}{emulates having real-time error information, as described in Section \ref{ssec:calibration_data}.}

To tackle the challenge of not having access to real-time calibration information, we explore several techniques for processing historical calibration data as outlined in Table~\ref{tab:calibration_set_classification}. They focus on the \twoq{} gate and readout error rates, as they are the main sources of noise that considerably affect circuit fidelity~\cite{wilson_just--time_2020, das_imitation_2023, murali_software_2020}. The experiments will compare the latest calibration data with different processing methods (See Sec.~\ref{ssec:calibration_data_sets}). \textcolor{black}{Fig.~\ref{fig:daily_line_perth_adder} demonstrates that the use of historical calibration data can reduce the impact of unexpected changes on the real device. Observe that the circuit fidelity when using TriQ-w-15 (\textcolor{black}{orange} line) does not experience a sharp decrease
on Day-6. However, on the bottom graph (Simulator), TriQ-w-15 shows a slight decrease of around 10\% compared to TriQ-lcd on Day-9 and Day-10. The same observation can also be seen from other circuits (Fig.~\ref{fig:summary_triq_sim_vs_real}), where historical calibration data has a smaller deviation (error bar) compared to the latest calibration data.}

Based on these experiments
, in which we use the noisy simulator to emulate real-time calibration data, we can conclude that \textbf{acquiring calibration data closer to real-time error conditions of the device provides the best overall circuit fidelity}, as expected\cite{wilson_just--time_2020, yeter-aydeniz_measuring_2023}. However, since this configuration can be time and resource-intensive and is not available in most cloud-based services, it is important to have an alternative that mitigates the performance loss. For this reason, we examined the impact of different historical calibration data models.

\subsection{Does using historical calibration data always lead to higher circuit performance (fidelity) than utilizing the latest calibration data? Does the size of the device impact on the circuit fidelity enhancement?}


The purpose of using historical calibration data is to compensate for any machine drift that may occur on the device. To further investigate this, we compute additional calibration processing methods according to the classification outlined in Table~\ref{tab:calibration_set_classification}. It is important to highlight that there may be multiple approaches to processing this information that can be explored in the future. For this work, the aim is to demonstrate that introducing historical error information in the compilation process might be beneficial. These proposed methods will be applied to the compilation techniques shown in Table~\ref{tab:technique_classification}. 


\begin{figure}[htbp]
\centering
\begin{subfigure}[b]{0.47\textwidth}
  \centering
  \includegraphics[width=.99\linewidth]{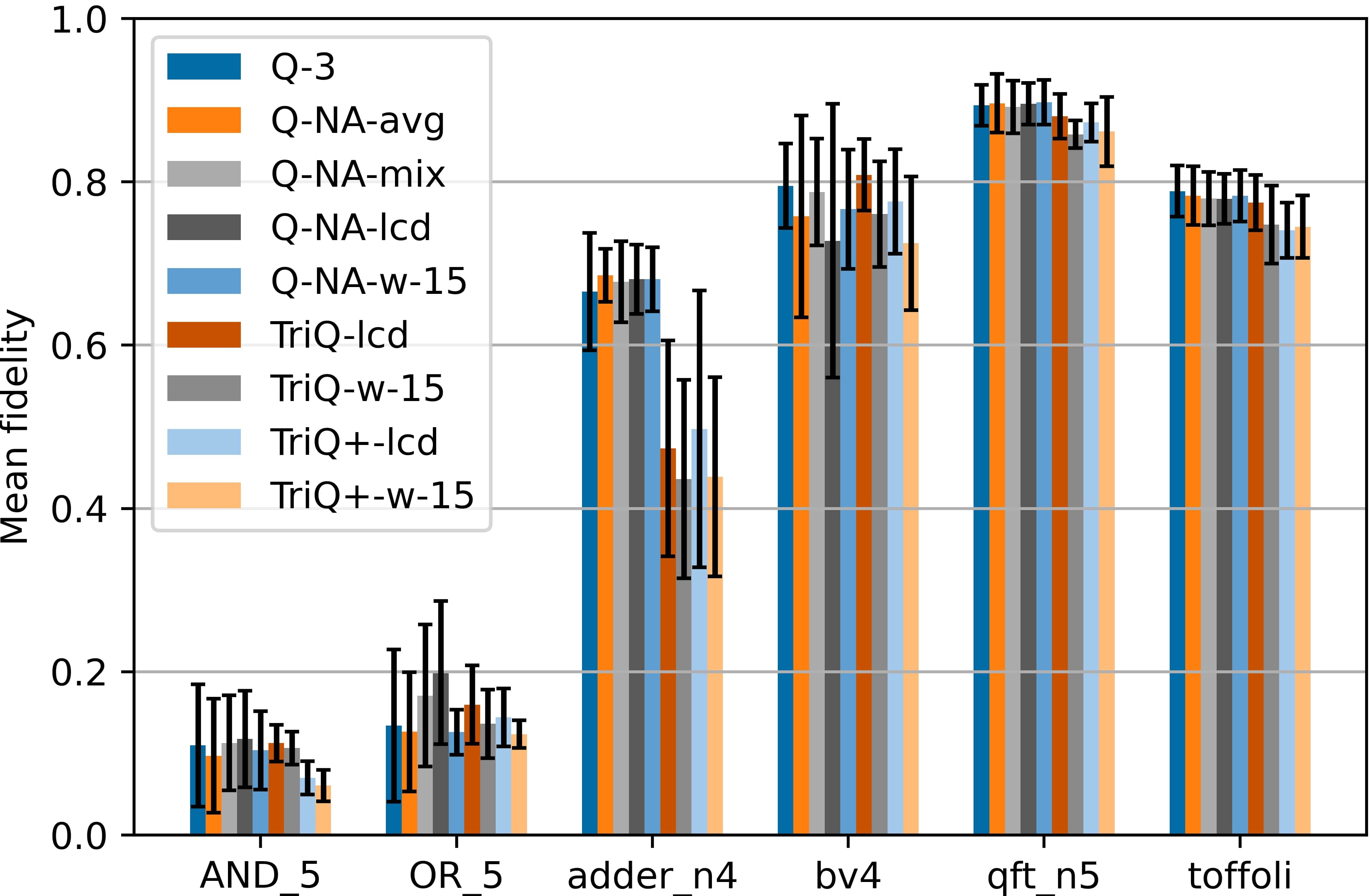}
  \caption{}
  \label{fig:summary_ibm_perth_real}
\end{subfigure}%
\hfill
\begin{subfigure}[b]{0.47\textwidth}
  \centering
  \includegraphics[width=.99\linewidth]{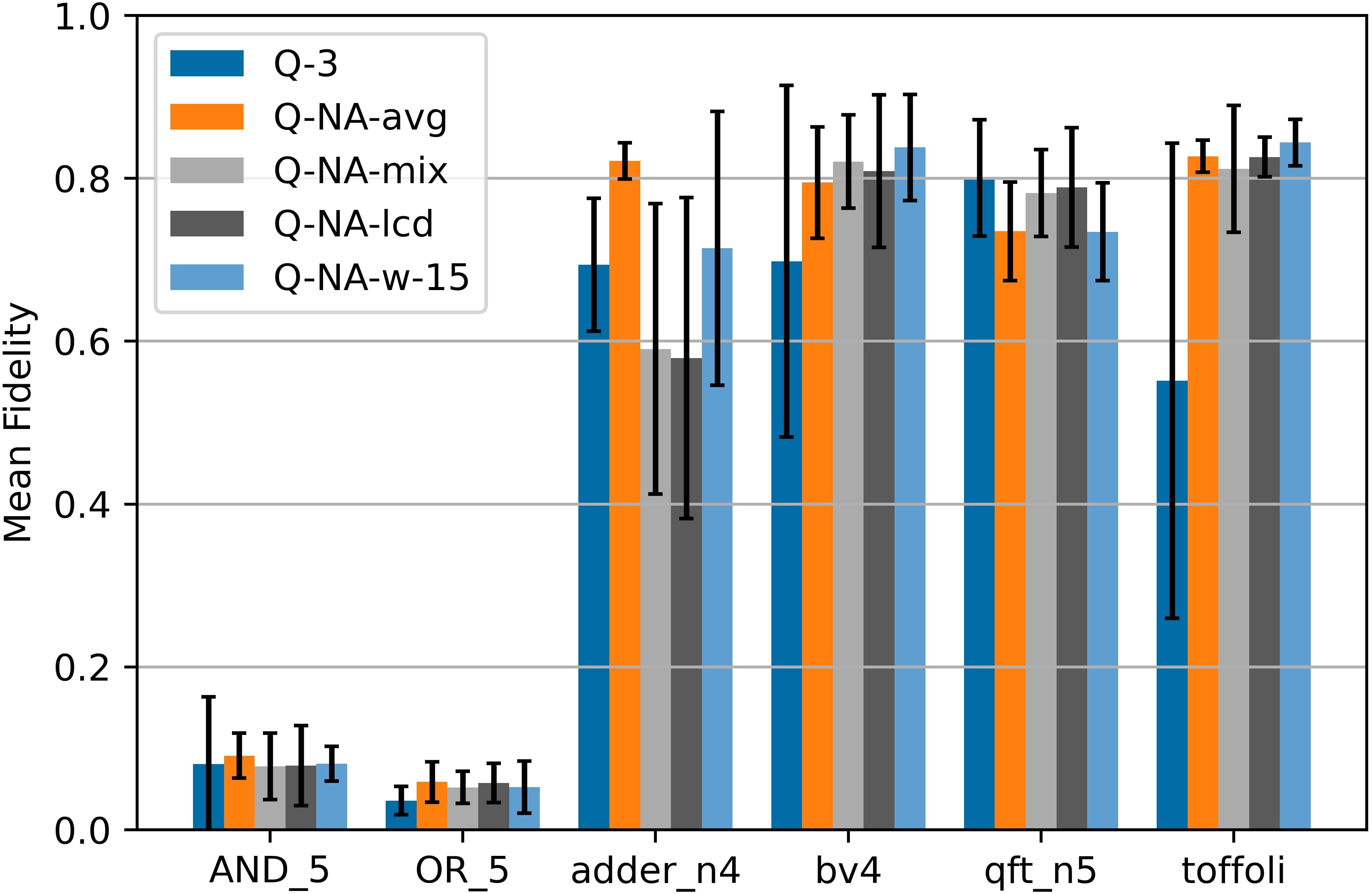}
  \caption{}
  \label{fig:summary_ibm_brisbane}
\end{subfigure}
\caption{\textcolor{black}{Algorithm fidelity when using noise-aware and non-noise-aware compilation techniques and different calibration process methods for (a) \perth{}, (b) \brisbane{}}.}
\vspace{-1.8em}
\label{fig:summary_perth_sim_vs_real}
\end{figure}


Fig.~\ref{fig:summary_ibm_perth_real} presents a summary of the experiments. It shows the advantage of noise-aware compilation techniques over non-noise-aware; it also compares several noise-aware methods when different historical calibration data is considered for both quantum processors. For example, Q-NA outperforms TriQ, particularly for the circuit adder\_n4, due to its more efficient routing technique compared to TriQ's reverse SWAP. However, with a circuit in which no qubit routing is needed (bv4), TriQ achieves the highest mean fidelity because it chooses higher fidelity qubits through the use of a reliability matrix in the initial qubit mapping stage. This kind of experiment indicates that there is no outperforming noise-aware solution, and all of them can be improved to increase fidelity in different quantum circuits.

Moreover, to assess the scalability of noise-aware compilation techniques to larger devices, we also conducted experiments on \brisbane{}, which has 127 qubits. We took into account the differences in qubit count and processor types (Falcon r5.11 with CNOT in \perth{}, Eagle r3 with ECR in \brisbane{}). Our findings align with ~\cite{sharma_noise-aware_2023}, indicating that larger devices offer more diverse mapping and routing configurations, enhancing the potential benefits of noise-aware compilation techniques. Fig.~\ref{fig:summary_ibm_brisbane} illustrates circuit fidelity improvements between 2\% and 100\% on \brisbane{}, with enhancements in five out of six algorithms. In contrast, \perth{} demonstrates more modest improvements (0.44\% to 48\%) in four out of six algorithms. \textbf{This highlights the impact of device size on noise-aware techniques and emphasizes their potential to enhance performance in larger quantum processors.} From the results, it can also be concluded that the choice of approach for processing the calibration data is important as it influences the magnitude of the fidelity improvement.

\begin{figure}[ht]
\centering
\begin{subfigure}[b]{0.47\textwidth}
  \centering
  \includegraphics[width=.96\linewidth]{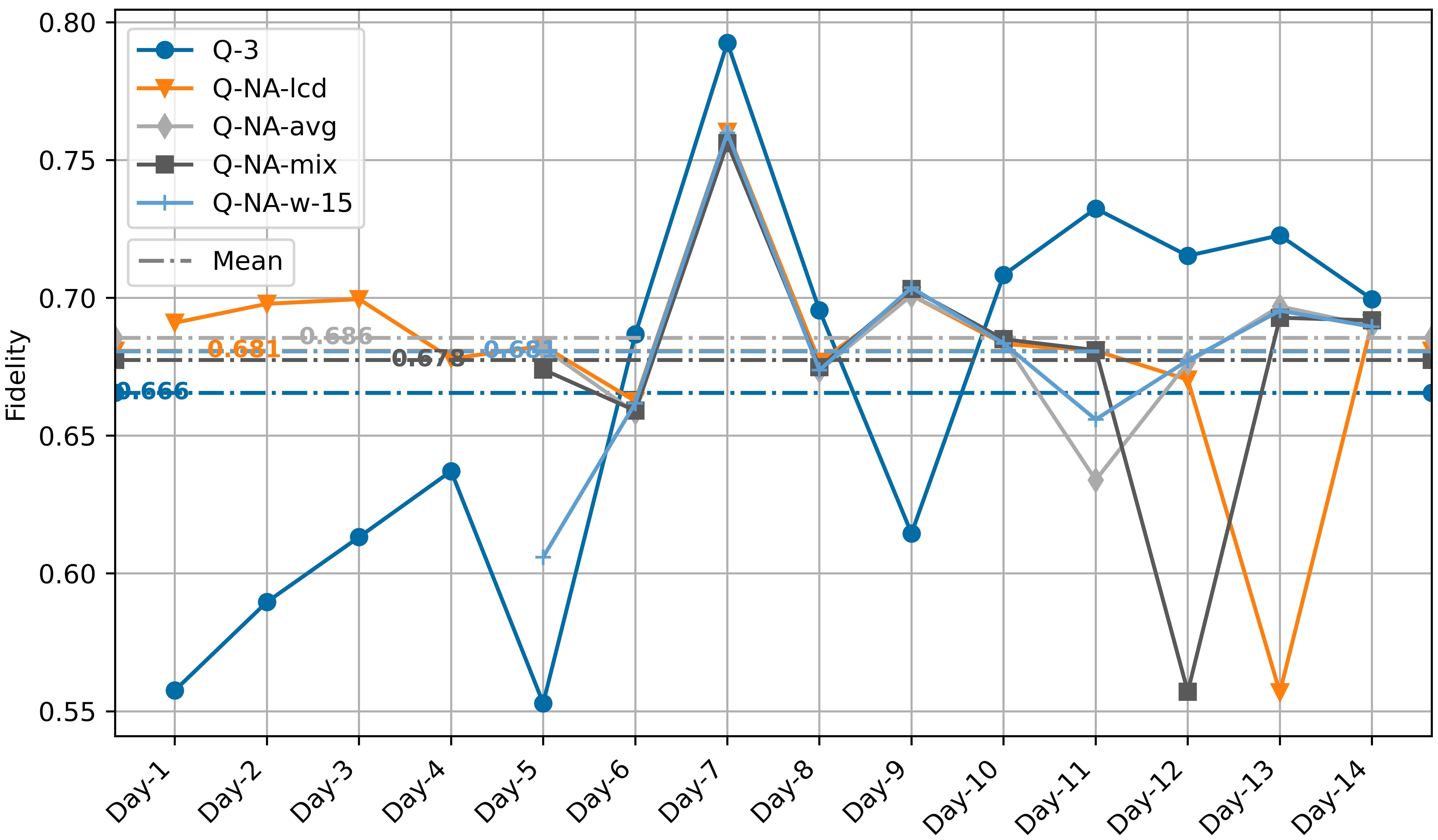}
  \caption{}
  \label{fig:line_adder_qiskit}
\end{subfigure}%
\hfill
\begin{subfigure}[b]{0.47\textwidth}
  \centering
  \includegraphics[width=.96\linewidth]{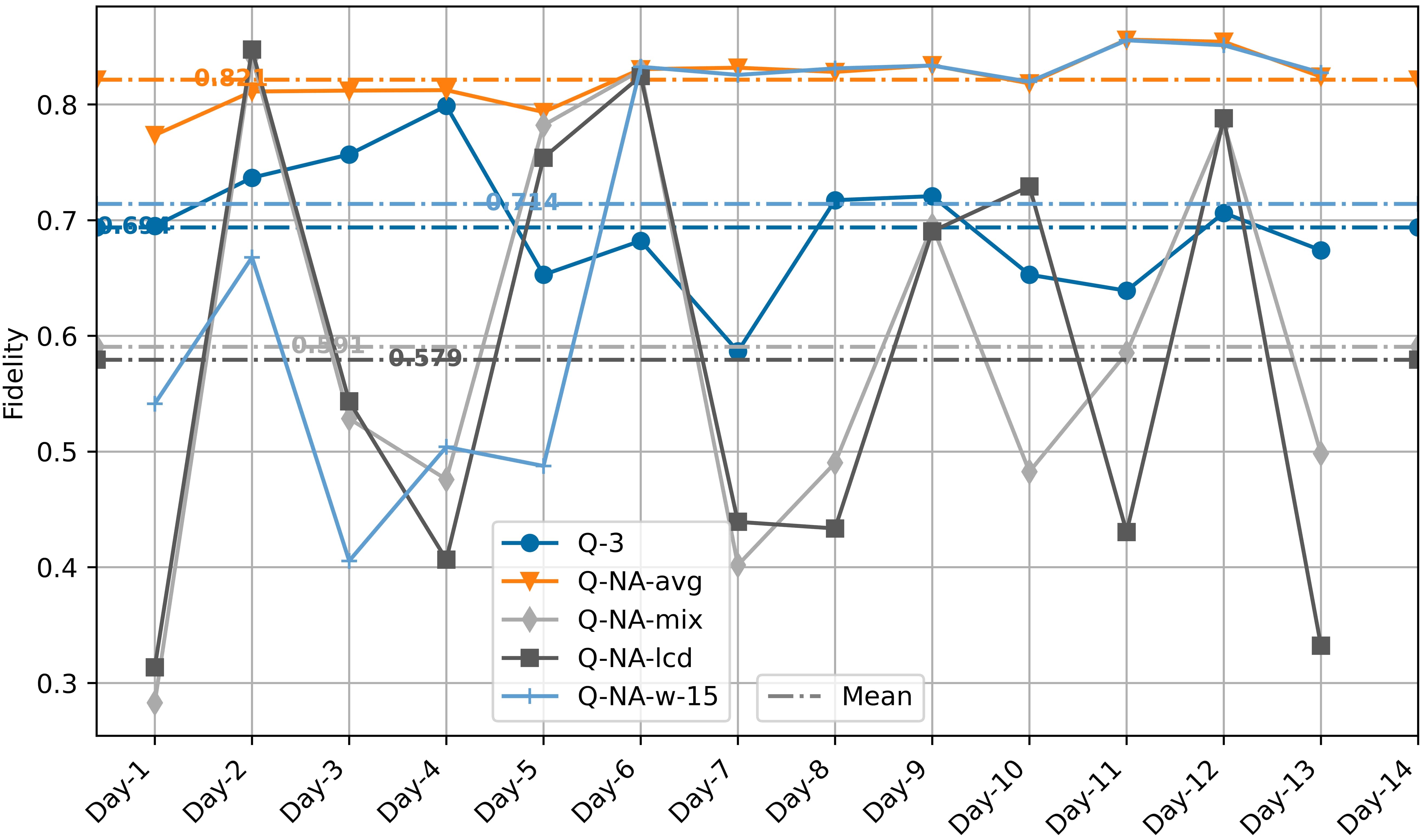}
  \caption{}
  \label{fig:daily_line_brisbane_fidelity}
\end{subfigure}
\caption{Daily fidelity of adder\_n4 circuit when executed on (a) \perth{} and (b) \brisbane{} using different noise-aware compilation techniques and historical calibration data processing methods. }
\vspace{-1.0em}
\label{fig:daily_line_perth_brisbane}
\end{figure}

Finally, we also performed a more detailed analysis of the daily fidelity obtained with each of the techniques for the different circuits from Table~\ref{tab:benchmark_list}. As an example, Fig.~\ref{fig:line_adder_qiskit} shows the daily fidelity of an adder\_n4 circuit with various calibration models on both \perth{} and \brisbane{} backends. Q-3 shows noticeable fluctuations, while Q-NA-lcd demonstrates stability, consistently approaching the mean. However, relying solely on the latest calibration data may expose vulnerabilities. For example, on Day-13, Q-NA-lcd experienced a significant drop in contrast to Q-NA-avg, which maintained fidelity. Results from \brisbane{}, which has more options for qubit mapping and routing, show increased variability. Fig.~\ref{fig:daily_line_brisbane_fidelity} highlights the benefits of noise-aware calibration, with configurations like Q-NA-avg exhibiting stability and the highest mean. So we can conclude that \textbf{in the absence of real-time calibration data, noise-aware compilation techniques that rely solely on the latest calibration data can result in greater deviations. In contrast, utilizing historical calibration data can prevent sharp deviations and result in a higher average fidelity}.


\subsection{What is the impact on the algorithm's fidelity as the circuit size increases?}


\begin{figure*}[t]
\centerline{\includegraphics[width=.90\linewidth]{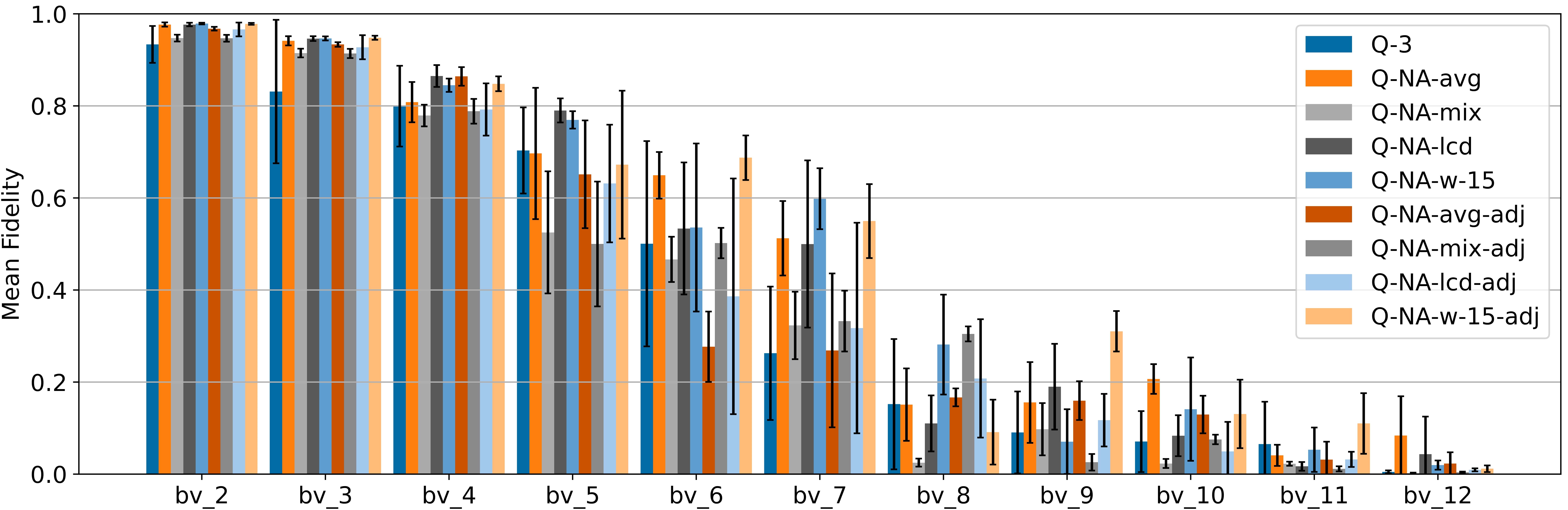}}

\caption{\textcolor{black}{Algorithm fidelity using different compilation techniques as circuit depth and width increases}. (run = 1, shots = 8192)}
\vspace{-1.5em}
\label{fig:brisbane_scale_bv}
\end{figure*}

Moving from our in-depth examination of small quantum circuits, we shift our focus to larger quantum algorithms (higher circuit width and depth).
\textcolor{black}{Fig.~\ref{fig:brisbane_scale_bv} demonstrates that using noise-aware compilation techniques yields greater benefits for larger circuits compared to the non-noise-aware ones, like Q-3.} For the experiments conducted on \brisbane{}, this improvement varies greatly, ranging from 4.83\% (bv\_2) to a significant 127.79\% (bv\_7) depending on the circuit and method of processing calibration data. A trend can be seen, indicating a decrease in fidelity as the circuit size increases. Nevertheless, certain methods of processing calibration data demonstrate improved performance on larger circuits. For instance, Q-NA-w-15-adj exhibits better fidelity in larger circuits despite its suboptimal performance in smaller circuits. \textbf{While noise-aware compilation techniques enhance fidelity as the circuit size increases, no technique consistently outperforms the others.}


\subsection{What is the correlation between circuit depth and the total number of \twoq{} gates with the algorithm's fidelity?}

    


Most of the compilation techniques aim to minimize circuit depth and/or the overhead of \twoq{} gates to mitigate issues related to decoherence and instability arising from \twoq{} gates~\cite{li_tackling_2019, liu_tackling_2023, mckinney_mirage_2023}. In this set of experiments, we explore the impact of circuit depth and the total number of \twoq{} gates (after compilation) on the circuit fidelity when using several noise-aware compilation techniques.

\begin{figure}[ht!]
    \centering
    \includegraphics[width=.99\linewidth]{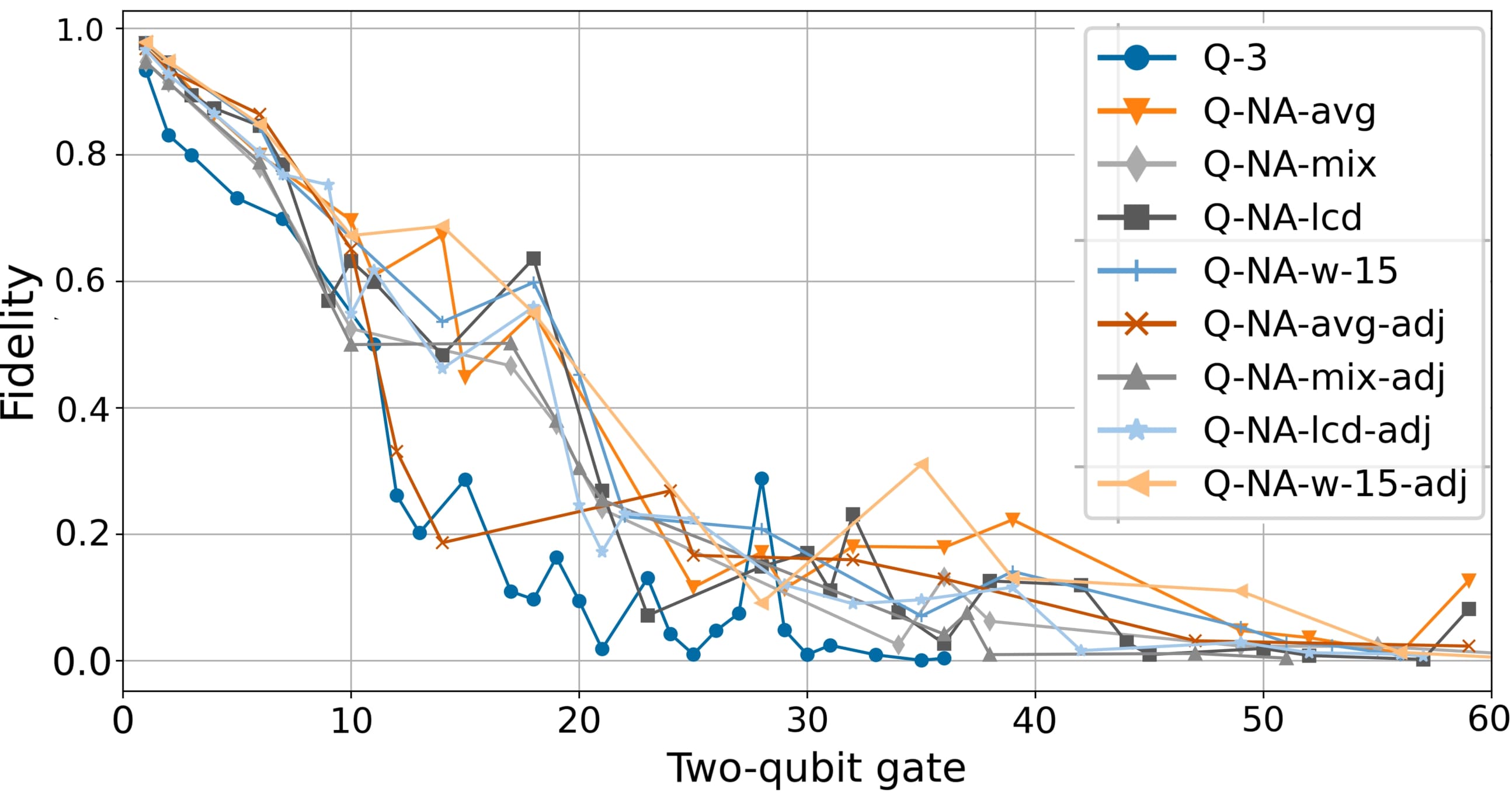}
    \caption{Impact of the \twoq{} gate on the BV fidelity when executed on \brisbane{} considering various calibration data processing methods. (run = 1, shots = 8192)}
    \vspace{-1.0em}
\label{fig:brisbane_scale_2q_bv}
\end{figure}

Fig.~\ref{fig:brisbane_scale_2q_bv} displays the total number of \twoq{} gates based on the circuits presented in Fig.~\ref{fig:brisbane_scale_bv}. The data indicates that, in general, as the total number of \twoq{} gates increases, the circuit fidelity decreases. However, Q-NA-w-15-adj and Q-NA-avg have the highest fidelity despite having a higher number of \twoq{} gates. It is suggested that \textbf{using historical calibration data in noise-aware compilation techniques may result in a higher number of \twoq{} gates, but it does not necessarily mean having lower circuit fidelity. In addition, as shown before, it exhibits higher performance than the latest calibration data methods}.

\begin{figure}[ht!]
    \centering
    \includegraphics[width=.99\linewidth]{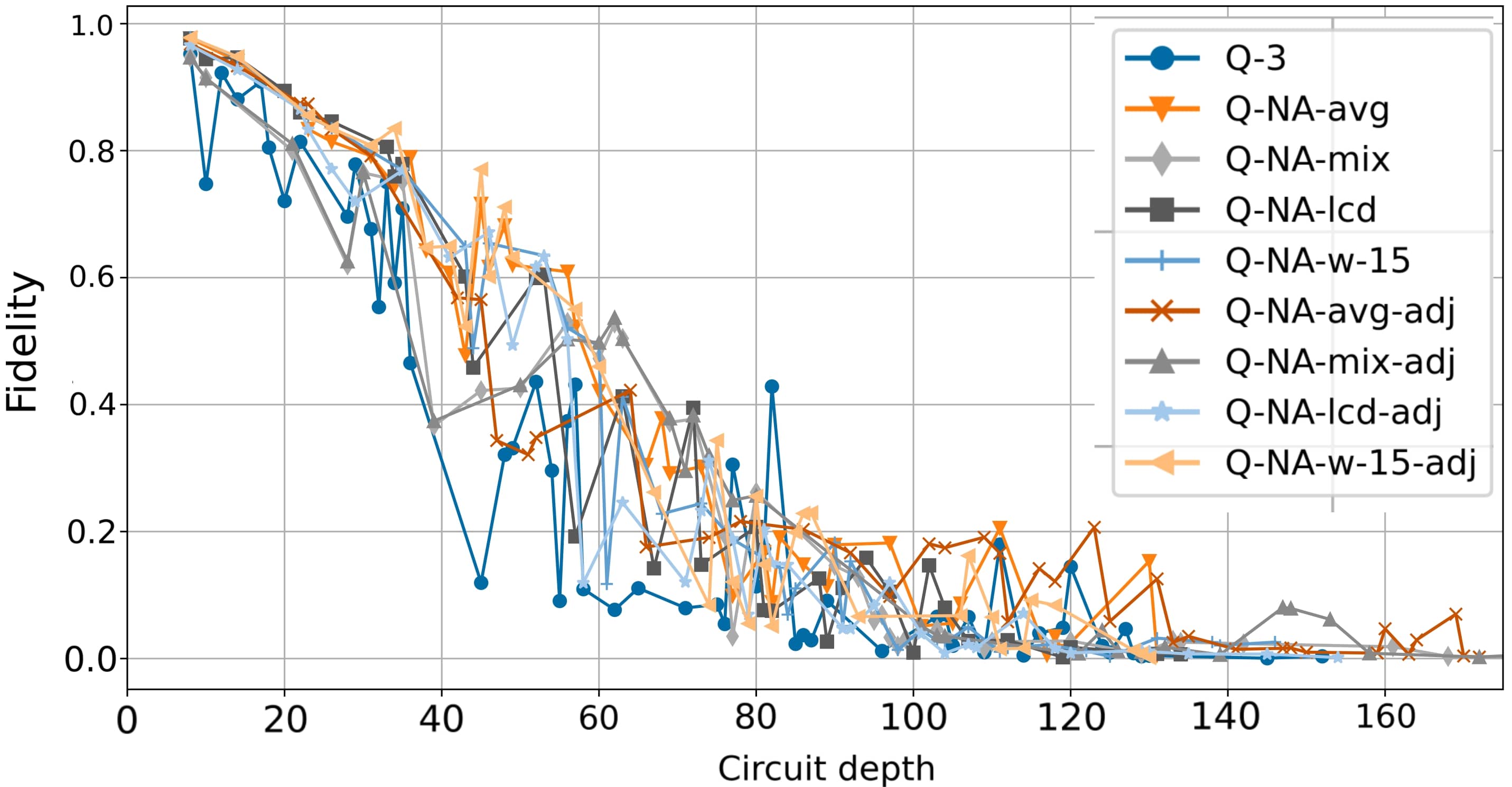}
    \caption{Impact of the circuit depth on the BV fidelity when executed on \brisbane{} considering various calibration data processing methods. (run = 1, shots = 8192)}
    \vspace{-1.0em}
\label{fig:brisbane_scale_depth_bv}
\end{figure}

\textcolor{black}{The relationship between circuit depth and fidelity is well-established: the higher the circuit depth, the lower the circuit fidelity is. However, as shown in Fig.~\ref{fig:brisbane_scale_depth_bv} (circuits from Fig.~\ref{fig:brisbane_scale_bv}), historical calibration data can lead to a high circuit depth but still show an increased fidelity. 
Similar to \twoq{} gates, it can also be observed that using historical calibration data may result in a higher circuit depth. Despite the increase in depth, the benefits of this approach outweigh them.}

Based on the previous two figures related to \twoq{} gates and circuit depth, we can conclude that \textbf{using only these metrics is insufficient to represent overall fidelity. }



\section{Conclusions} \label{sec:conclusion}

Noise-aware circuit compilation techniques are promising to further improve the overall circuit fidelity in NISQ devices as a result of selecting qubits with small readout errors for the initial qubit mapping and reliable paths for the qubit routing stage. In this paper, we have developed a framework that supports running multiple circuits with various calibration techniques under the same conditions within a single quantum job on a vendor processor. This framework also allows for the execution of circuits on both noisy simulators and real devices. With the help of this framework, we examine the benefits of using noise-aware compilation techniques compared to non-noise-aware techniques. The \textcolor{black}{study found} that the overall fidelity achieved with noise-aware compilation techniques is greater when the calibration data is closer to real-time data compared to non-noise-aware techniques. When comparing the results of the noisy simulator and the real device, it was observed that noise-aware compilation techniques relying solely on the latest calibration data can \textcolor{black}{yield
} greater deviation in the absence of real-time calibration data. This led to the definition of different approaches for processing calibration data. The experiments demonstrated that incorporating historical calibration data can prevent deviation and lead to a more accurate average.  Additionally, the results indicated that the use of high-fidelity qubits in \twoq{} gates and measurements is more crucial than the number of circuit depth and \twoq{} gates employed. Therefore, increasing the depth and the number of gates may still \textcolor{black}{provide} a better final fidelity when using high-fidelity qubits. It can be concluded that using only the metrics of circuit depth and \twoq{} gates is insufficient to represent overall fidelity.

All experiments compared two different topologies in \perth{} (7 qubits) and \brisbane{} (127 qubits). The results demonstrate that for noise-aware compilation techniques to work efficiently, a larger device is needed as it provides more options and chances to increase overall fidelity. Based on the experiments conducted, it was concluded that no approach for processing calibration data was found to be a universal strategy among the noise-aware techniques that is optimum. The degree of improvement provided varied depending on the specificities of the quantum circuit being run. 

In future work, we will investigate how to characterize calibration data in a more sophisticated manner and combine error information with QEM. \textcolor{black}{Additionally, we will examine the impact of applying these techniques to other quantum computing platforms.}

\section*{Acknowledgment}

This work is supported by the QuantERA grant EQUIP with the grant numbers PCI2022-133004 and PCI2022-132922, funded by Agencia Estatal de Investigación, Ministerio de Ciencia e Innovación, Gobierno de España, MCIN/AEI/10.13039/501100011033, and by the European Union “NextGenerationEU/PRTR”. CGA acknowledges funding from the Spanish Ministry of Science, Innovation and Universities through the Beatriz Galindo program 2020 (BG20-00023) and from the  European ERDF under grant PID2021-123627OB-C51. We acknowledge the use of IBM Quantum services for this work. The views expressed are those of the authors and do not reflect the official policy or position of IBM or the IBM Quantum team.
\bibliographystyle{ieeetr}
\bibliography{calibration}



\end{document}